\documentclass[aps,prfluids,preprint,superscriptaddress]{revtex4-1}

\usepackage[utf8]{inputenc}
\usepackage{amssymb,amsmath}
    \makeatletter
    \renewcommand*\env@matrix[1][*\c@MaxMatrixCols c]{%
      \hskip -\arraycolsep
      \let\@ifnextchar\new@ifnextchar
      \array{#1}}
    \makeatother
\usepackage{array}                      
\usepackage[titletoc]{appendix}
\usepackage{bm}                         
\usepackage{booktabs}
\usepackage[margin=2.5cm]{geometry}
\usepackage{graphicx}
\usepackage{multirow}                   
\usepackage{setspace}
\usepackage[caption=false,position=top]{subfig}
\usepackage{lineno}
\usepackage{listings}
\usepackage{diagbox}
\usepackage{natbib}

\usepackage{lineno}

\renewcommand{\epsilon}{\varepsilon}

\newcommand{\MATLAB}{MATLAB\textsuperscript{\textregistered}~}

\graphicspath{{FIGS/}}

\begin{document}

\title{Efficient Implementation of Elastohydrodynamics via Integral Operators}

\author{A.\,L.\ Hall-McNair}
\affiliation{School of Mathematics, University of Birmingham, \\ Edgbaston, Birmingham, UK, B15 2TT}
\author{T.\,D.~Montenegro-Johnson}
\affiliation{School of Mathematics, University of Birmingham, \\ Edgbaston, Birmingham, UK, B15 2TT}
\author{H.\, Gad\^{e}lha}
\affiliation{Faculty of Engineering, University of Bristol, \\ Bristol, UK, BS8 1UB}
\author{D.\,J.\ Smith}
\affiliation{School of Mathematics, University of Birmingham, \\ Edgbaston, Birmingham, UK, B15 2TT}
\author{M.\,T.\ Gallagher \footnote{Email address for correspondence: \texttt{m.t.gallagher@bham.ac.uk}}}
\affiliation{School of Mathematics, University of Birmingham, \\ Edgbaston, Birmingham, UK, B15 2TT}

%

\begin{abstract}
    The dynamics of geometrically non-linear flexible filaments play an important role in a host of biological processes, from flagella-driven cell transport to the polymeric structure of complex fluids. Such problems have historically been computationally expensive due to numerical stiffness associated with the inextensibility constraint, as well as the often non-trivial boundary conditions on the governing high-order PDEs. Formulating the problem for the evolving shape of a filament via an integral equation in the tangent angle has recently been found to greatly alleviate this numerical stiffness. The contribution of the present manuscript is to enable the simulation of non-local interactions of multiple filaments in a computationally efficient manner using the method of regularized stokeslets within this framework. The proposed method is benchmarked against a non-local bead and link model, and recent code utilizing a local drag velocity law. Systems of multiple filaments (1) in a background fluid flow, (2) under a constant body force, and (3) undergoing active self-motility are modeled efficiently. Buckling instabilities are analyzed by examining the evolving filament curvature, as well as by coarse-graining the body frame tangent angles using a Chebyshev approximation for various choices of the relevant non-dimensional parameters. From these experiments, insight is gained into how filament-filament interactions can promote buckling, and further reveal the complex fluid dynamics resulting from arrays of these interacting fibers. By examining active moment-driven filaments, we investigate the speed of worm- and sperm-like swimmers for different governing parameters. The \MATLAB implementation is made available as an open-source library, enabling flexible extension for alternate discretizations and different surrounding flows.
\end{abstract}

\maketitle

\section{Introduction} \label{sec:introduction}

Flexible filaments are ubiquitous in the natural world, and thus a clear understanding of their behavior is paramount in many biological problems. Models for simulating the dynamics of these filaments, while plentiful, have historically been mathematically complex and numerically expensive, with even simple computational experiments taking in the order of hours or even days to solve (see Moreau \textit{et al.} for detailed benchmarking \cite{moreau2018asymptotic}).

Micro-scale filament problems have been previously tackled using other modeling approaches which can be broadly separated into those based upon (a) a nonlinear PDE in the filament position (such as the method by Schoeller \textit{et al.} \cite{schoeller2019method}), or (b) a discretization into simpler elements, such as beads with connecting springs \cite{jayaraman2012autonomous} or interlocking gears \cite{delmotte2015general}. For category (a), inextensibility is enforced using Lagrange multipliers of tension, which are often costly to compute. For the discrete approaches in category (b), other ways of enforcing this condition are used. For example, the bead model of Jayaraman \textit{et al.} \cite{jayaraman2012autonomous} prescribes large spring constants between each bead, contributing to the numerical stiffness of the system. Equivalently, the gears model of Delmotte \textit{et al.} \cite{delmotte2015general} imposes a non-holonomic constraint to ensure non-penetrability between adjacent beads, but as a result requires large numbers of points to represent a single filament.

A recent promising development via Moreau \textit{et al.} \cite{moreau2018asymptotic}, referred to as \textit{coarse graining}, is based on reformulating the problem via an integral equation with the filament tangent angle as the dependent variable. The method, initially developed using a local hydrodynamic drag law, provides an efficient framework for simulating non-interacting filament dynamics. This approach builds upon the early studies of Brokaw \cite{brokaw1971bend,brokaw1972computer} and Hines \& Blum \cite{hines1978bend}, and contrasts with Cartesian formulations \cite{gadelha2010nonlinear,tornberg2004simulating}.

The contribution of the present manuscript is to enable efficient and accurate simulation of multiple, non-locally interacting, passive and active filaments in ambient flows by incorporating recent developments in the regularized stokeslet method \cite{cortez2018regularized,gallagher2018meshfree} with the integral formulation in terms of the tangent angle of Moreau \textit{et al.} \cite{moreau2018asymptotic}.

The potential applications for a fast and accurate filament modeling framework are numerous. There has long been interest in understanding the mechanics and regulation of sperm flagellar movement, in particular problems relating to: understanding the mechanical structure and motor regulation \cite{brokaw1971bend,lindemann1994model,guo2018bistability,oriola2017nonlinear}, investigating the response of the flagellar beat to its rheological environment \cite{huang2018hydrodynamic,gadelha2013counterbend,smith2009bend}, understanding the dynamics of sperm due to surrounding solid walls \cite{montenegro2015spermatozoa,denissenko2012human}, and studying the effect of viscosity on sperm swimming \cite{woolley2001study}. For a detailed review surrounding the importance of the sperm flagellum see Gallagher \textit{et al.} \cite{gallagher2018casa}. Furthermore, such a method could be used to investigate phenomena associated with epithelial cilia driven flows such as: cilia waveform modulation by length \cite{brokaw1971bend}, the effects of flow induced by cilia on embryonic development \cite{montenegro2012modelling}, studying the physical limits of flow sensing \cite{ferreira2017physical}, and investigating the mechanical structure of the axoneme in cilia \cite{omori2017nodal}. Another avenue of active-filament research to which the proposed framework could be applied is magnetic swimmers \cite{gadelha2013optimal}. These models have wider relevance in the field of synthetic biology, with particular application to microscopic bacteriophage-based fibre sensors \cite{pacheco2011detection,lobo2015direct,gallagher2017model} and flexible filament microbots \cite{montenegro2018microtransformers}. The proposed framework could be used to further investigate the dynamics of bundles of filaments \cite{coy2017counterbend}, and additionally has applications in the multi-scale studies of complex polymeric fluids, and flagellar movement through them \cite{yang2017dynamics,wrobel2016enhanced}.

We will extend the framework introduced by Moreau \textit{et al.} \cite{moreau2018asymptotic}, augmenting and reformatting their formulation with the method of regularized stokeslets of Cortez \textit{et al.}~\cite{cortez2001method,cortez2005method}. These methods have proven to be accurate and effective in modeling the hydrodynamics in various multiple-fiber scenarios \cite{stein2019coarse,olson2015hydrodynamic}. The method of regularized stokeslets enables the modeling of non-local hydrodynamics within and between filaments, and between filaments and surrounding structures. The method is implemented in a numerically efficient manner, retaining the computational economy and low hardware requirements inherited from the Moreau \textit{et al.} formulation.

The structure of this manuscript is as follows: in Sec.~\ref{sec:model-formulation}, the Elastohydrodynamic Integral Formulation (EIF) for a single filament is proposed. In Sec.~\ref{sec:single-multi-filament-problems}, alterations to the EIF for various single- and multi-filament scenarios are presented.  Verification and benchmarking of the method is given in Sec.~\ref{sec:verification}. Simulation results of the problems formulated in Sec.~\ref{sec:single-multi-filament-problems} are then presented in Sec.~\ref{sec:results}, followed by discussion of the results and of further possible applications in Sec.~\ref{sec:discussion}. The MATLAB\textsuperscript{\textregistered} code for the methods described within this report are provided in the associated GitLab repository, available at {\texttt{gitlab.com/atticushm/eif}}.

\section{Model formulation} \label{sec:model-formulation}

The dynamics of elastic filaments in Stokes flow will be modeled by constructing integral operator formulations of the governing fluid and elastodynamic equations. Each filament is described by the time-evolving tangent angle $ \theta(s,t) $ along its arclength. Taking $\bm{X}_0\left(t\right)$ to be the leading point at time $t$, the filament geometry is then given by
\begin{linenomath*} \begin{equation}
    \bm{X}(s,t) = \bm{X}_0(t)+\int_0^s [\cos\theta(s',t),\sin\theta(s',t),0]^T \,ds', \label{eq:kinematic}
\end{equation} \end{linenomath*}
where $s\in[0,1]$ and $\theta\in[-\pi,\pi)$ are the dimensionless arclength (scaled by the filament length $ L $) and tangent angle respectively. The velocity of a point on the filament is given by differentiating Eq.~\eqref{eq:kinematic} with respect to dimensionless time, scaled by $ \tau $. The behavior of planar filaments in a Newtonian fluid is considered, which can be described by the three-dimensional dimensionless Stokes flow equations,
\begin{linenomath*} \begin{equation}
    \bm{0}=-\bm{\nabla}p+\nabla^2\bm{u} + \bm{F}, \quad \nabla\cdot\bm{u}=0,\label{eq:stokes}
\end{equation} \end{linenomath*}
where \(\bm{u}(\bm{x},t)\) is fluid velocity, \(p(\bm{x},t)\) is pressure, and $ \bm{F}(\bm{x},t) $ is a force exerted by the body onto the fluid. As shown by Cortez \cite{cortez2005method}, an approximate solution to Eq.~\eqref{eq:stokes} is given by the regularized stokeslet integral,
\begin{linenomath*} \begin{equation}
    u_j(\bm{x},t) = \frac{1}{8\pi}\int_0^1 S_{jk}^\epsilon\left(\bm{x},\bm{X}(s',t)\right)f_k(s',t)\,ds' + \mathcal{O}\left(\epsilon^r\right), \label{eq:sbt}
\end{equation} \end{linenomath*}
where \(\bm{f}(s,t)\) is the force per unit length exerted by the filament on the fluid, non-dimensionalized with the scaling $ \mu L/\tau $ for a given fluid dynamic viscosity $ \mu $, and $ \bm{X}(s,t) $ denotes the filament position as a function of arclength and time. The error arises from the regularization of the stokeslet, for a chosen regularization parameter $ 0<\varepsilon\ll 1 $, where $ r=1 $ or $ 2 $ in the near- and far-field respectively. The combined process of solving Eq.~\eqref{eq:sbt} for the unknown force densities $ \bm{f}(s,t) $ will result in errors of order $ \varepsilon $ everywhere \cite{martin2019use}.
Summation convention dictates that repeated indices are summed over and unrepeated indices range over $\{1,2,3\}$. The kernel of the integral in Eq.~\eqref{eq:sbt} is known as the regularized stokeslet, defined
\begin{linenomath*} \begin{equation}
	S_{jk}^{\varepsilon}\left(\bm{x},\bm{X}\right) = \delta_{jk}\frac{|\bm{x}-\bm{X}|+2\varepsilon^2}{\left(|\bm{x}-\bm{X}|^2 + \varepsilon^2\right)^{3/2}} + \frac{\left(x_j-X_j\right)\left(x_k-X_k\right)}{\left(|\bm{x}-\bm{X}|^2 + \varepsilon^2\right)^{3/2}}.
\end{equation} \end{linenomath*}
Applying the no-slip condition $ \bm{u}(\bm{X}(s,t),t)=\partial_t\,\bm{X}(s,t) $ on the boundary of the filament yields the regularized stokeslet integral equation
\begin{linenomath*} \begin{equation}
    \partial_t \,X_j(s,t) = \frac{1}{8\pi}\int_0^1 S_{jk}^\epsilon\left(\bm{X}(s,t),\bm{X}(s',t)\right)f_k(s',t)\,ds', \quad 0\leqslant s \leqslant 1, \quad t\geqslant 0. \label{eq:conthydro}
\end{equation} \end{linenomath*}
Inertialess dynamics requires that the filament is force and moment free at each instant, giving
\begin{linenomath*}\begin{align}
    \int_0^1-\bm{f}(s,t)\,ds &       			              = \bm{0},\label{eq:totalforcebalance} \\
    \int_0^1\bm{X}(s,t) \times\left(-\bm{f}(s,t)\right)\,ds &  = \bm{0}.\label{eq:totalmomentbalance}
\end{align}\end{linenomath*}
The integral formulation for the hydrodynamic problem thus comprises Eqs.~\eqref{eq:conthydro}, \eqref{eq:totalforcebalance} and \eqref{eq:totalmomentbalance}. Next, the elastic behavior of the filaments is considered.

The integral operator for the elastodynamic behavior is formulated by considering the force and moment balance over an infinitesimal segment of a filament. We denote by \(\bm{N}(s,t)\) and \(M(s,t)\) the contact force and contact moment respectively exerted by a distal segment of filament \([s,1]\) on a proximal segment \([0,s)\). Constitutively linear elasticity and bending in the $ xy $-plane implies that the dimensional contact moment relates to the curvature via
\begin{linenomath*} \begin{equation}
    M(s,t)=EI\,\partial_{s}\,\theta(s,t),\label{eq:constitutive}
\end{equation} \end{linenomath*}
where $ EI $ is the bending modulus of the filament. The fluid dynamic force density \(\bm{f}(s,t)\) and \(\bm{N}(s,t)\) are related by
\begin{linenomath*} \begin{equation}
    \partial_s\,\bm{N}(s,t) = \bm{f}(s,t). \label{eq:ehdforcebalance}
\end{equation} \end{linenomath*}
Under the assumption of free boundary conditions, the contact force and moment are zero at each end of a filament. Moment balance over the infinitesimal segment reveals that the contact force and moment are related as
\begin{linenomath*} \begin{equation}
    0=\partial_s\, M(s,t) + \bm{e}_3\cdot\partial_s\bm{X}(s,t)\times\bm{N}(s,t). \label{eq:momentdensity}
\end{equation} \end{linenomath*}
Integration of Eq.~\eqref{eq:momentdensity} over a distal segment $[s,1]$ yields, on application of the moment-free boundary condition and Eq.~\eqref{eq:ehdforcebalance},
\begin{linenomath*} \begin{equation}
    0 = -M(s,t) + \bm{e}_3\cdot\left[\big(\bm{X}(s',t)-\bm{X}(s,t)\big)\Big|_s^1-\int_s^1(\bm{X}(s',t)-\bm{X}(s,t))\times\bm{f}(s',t)\,ds'\right].
\end{equation} \end{linenomath*}
Application of the the force-free condition \(\bm{N}(1,t)=0\), the constitutive Eq.~\eqref{eq:constitutive}, and the choice of time scale $ \tau = \mu L^4/EI $, gives the non-dimensionalized integral formulation for the filament tangent angle evolution
\begin{linenomath*} \begin{equation}
    \partial_s\,\theta(s,t) = -\bm{e}_3\cdot\int_s^1(\bm{X}(s',t)-\bm{X}(s,t))\times\bm{f}(s',t)\,ds', \quad 0\leqslant s \leqslant 1, \quad t\geqslant 0. \label{eq:contelasto}
\end{equation} \end{linenomath*}
Combining Eqs.~\eqref{eq:kinematic}, \eqref{eq:conthydro}, \eqref{eq:totalforcebalance}, \eqref{eq:totalmomentbalance}, and \eqref{eq:contelasto}, we obtain the elastohydrodynamic integral formulation (EIF) for the evolution of a filament in Stokes flow
\begin{linenomath*} \begin{align}
    \partial_t\, X_j(s,t) &             = \frac{1}{8\pi}\int_0^1 S_{jk}^\epsilon\left(\bm{X}(s,t),\bm{X}(s',t)\right)f_k(s',t)\,ds',  \label{eq:fluidIE}\\
    \bm{0} &                            = \int_0^1-\bm{f}(s,t)\,ds, \\
    \bm{0} &                            = \int_0^1\bm{X}(s,t) \times\left( -\bm{f}(s,t)\right)\,ds, \\
    \partial_s\,\theta(s,t) &           = -\bm{e}_3\cdot\int_s^1(\bm{X}(s',t)-\bm{X}(s,t))\times\bm{f}(s',t)\,ds',  \quad 0 < s \leqslant 1, \quad t\geqslant 0, \label{eq:elastEQ}\\
    \mbox{where} \quad \bm{X}(s,t) &    = \bm{X}_0(t)+\int_0^s [\cos\theta(s',t),\sin\theta(s',t),0]^T \,ds', \\
    \mbox{and} \quad
    \partial_t\,\bm{X}(s,t) &           = \partial_t\,\bm{X}_0(t)+\int_0^s \partial_t\,\theta(s',t) [-\sin\theta(s',t),\cos\theta(s',t),0]^T \,ds',
\end{align} \end{linenomath*}
with unknowns \(\bm{X}_0(t)\), \(\theta(s,t)\) and \(\bm{f}(s,t)\). In solving the EIF for a particular physical problem, we must specify the initial position $\bm{X}_0(0)$ and tangent angle $\theta(s,0)$.

\subsection{Spatial discretization of the EIF} \label{subsec:semi-disc-of-eif}

\begin{figure}
	\centering
	\includegraphics[width=0.8\linewidth]{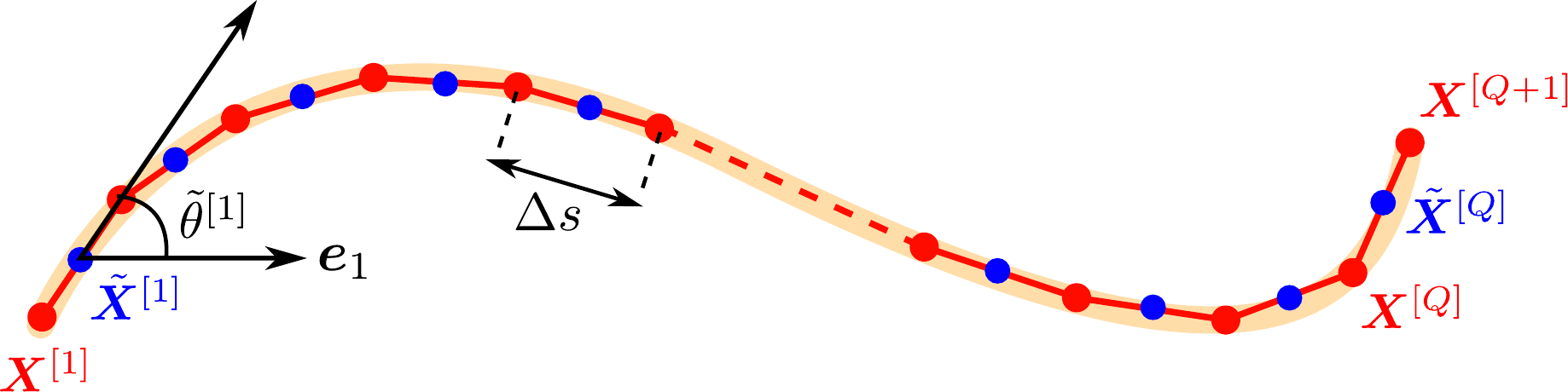}
	\vspace{0.2cm}
	\caption{
		Schematic illustrating the geometric discretization used in the EIF to model a continuous filament. The arclength is split into $ Q $ segments of equal length $ \Delta s $, with tangent angles $ \tilde{\theta}^{\,[n]} $ for $ n=1,\dots,Q $. Segment midpoints are represented by blue nodes, and segment endpoints are in red.
	}
	\label{figure1}
\end{figure}

To solve the EIF, we spatially discretize filaments to obtain a set of integro-differential equations which can be numerically evaluated. Dividing the arclength domain into \(Q\) segments of equal length \(\Delta s=1/Q\), the positions of the resulting segment endpoints are denoted as
\begin{linenomath*} \begin{equation}
    \bm{X}^{[n]}(t):=\bm{X}\left((n-1)\Delta s,t\right), \quad n=1,\ldots,Q+1.
\end{equation} \end{linenomath*}
The angle connecting \(\bm{X}^{[n]}(t)\) to \(\bm{X}^{[n+1]}(t)\) approximating \(\theta\big((n-1)\Delta s,t\big)\) is denoted \(\tilde{\theta}^{\,[n]}(t)\), for \(n=1,\ldots,Q\). The positions of the endpoints are given in terms of the initial point \(\bm{X}^{[1]}(t)\) and discretized tangent angle \(\tilde{\theta}^{\,[n]}(t)\) as
\begin{linenomath*} \begin{equation}
    \bm{X}^{[m+1]}(t)=\bm{X}^{[1]}(t)+\sum_{n=1}^m \Delta s \left[\cos\tilde{\theta}^{\,[n]}(t),\sin\tilde{\theta}^{\,[n]}(t),0\right]^T, \quad m=1,\ldots, Q, \label{eq:kine-shape}
\end{equation} \end{linenomath*}
with the segment midpoints
\begin{linenomath*} \begin{equation}
	\tilde{\bm{X}}^{[m]}(t) = \bm{X}^{[1]}(t)+\sum_{n=1}^{m-1} \Delta s \left[\cos\tilde{\theta}^{\,[n]}(t),\sin\tilde{\theta}^{\,[n]}(t),0\right]^T+\frac{\Delta s}{2} \left[\cos\tilde{\theta}^{\,[m]}(t),\sin\tilde{\theta}^{\,[m]}(t),0\right]^T, \label{eq:midpoints}
\end{equation} \end{linenomath*}
for each $m=1,\dots, Q$. An illustration of this discretization is displayed in Fig.~\ref{figure1}. Differentiating Eq.~\eqref{eq:kine-shape} with respect to time yields the kinematic equation for the segment velocities.

For the fluid dynamics, rather than using a conventional (and potentially expensive), quadrature rule for evaluating the rapidly-varying kernel $S_{jk}^\epsilon $, we employ the method of Smith \cite{smith2009boundary}. By approximating the force density in Eq.~\eqref{eq:fluidIE} as piecewise constant along each segment, the kernel can be analytically integrated, reducing the level of quadrature needed to evaluate the slowly-varying force density (for higher-order force discretizations see Cortez~\cite{cortez2018regularized}). Writing
\begin{linenomath*} \begin{equation}
    \bm{f}(s,t)\approx\tilde{\bm{f}}^{\,[m]}(t)=\bm{f}(\tilde{s}^{\,[m]},t), \qquad (m-1)\Delta s\leqslant s < m\Delta s, \label{eq:fdisc}
\end{equation} \end{linenomath*}
where $ \tilde{s}^{\,[m]} $ denotes the arclength at the midpoint of the $ m $\textsuperscript{th} segment, with a piecewise linear discretization for the filament,
\begin{linenomath*} \begin{equation}
    \bm{X}(s,t)\approx\tilde{\bm{X}}^{[m]}(t)+\left(s-\tilde{s}^{\,[m]}\right)\Big[\cos\theta\left(\tilde{s}^{\,[m]},t\right),\,\sin\theta\left(\tilde{s}^{\,[m]},t\right),\,0\Big]^T,
\end{equation} \end{linenomath*}
we obtain the spatially discrete equation,
\begin{linenomath*}
	\begin{align}
    &\partial_t\, \tilde{X}_j^{[m]}(t) = \frac{1}{8\pi}\sum_{n=1}^Q \tilde{f}_k^{\,[n]}(t)\nonumber \\
    & \qquad\cdot\int_{(n-1)\Delta s}^{n\Delta s} S_{jk}^\epsilon\left(\tilde{\bm{X}}^{[m]},\tilde{\bm{X}}^{[n]}(t)+\left(s'-\tilde{s}^{\,[n]}\right)\left[\cos\theta(\tilde{s}^{\,[n]},t),\,\sin\theta(\tilde{s}^{\,[n]},t),\,0\right]^T \right) ds'. \label{eq:fluidIE2}
\end{align}
\end{linenomath*}
The integral in Eq.~\eqref{eq:fluidIE2} can be calculated exactly by transforming to a local coordinate system in which one axis is aligned with the segment tangent (details are given in App.~B, Eqs.~B1--B3 of \cite{smith2009boundary}, although note that the simplified form for the near-field integral in Eq.~B4 contains a typographical error in the $\delta s/\epsilon$ fraction, which is upside-down). For brevity we denote the integral in Eq.~\eqref{eq:fluidIE2} as $I_{jk}^{\,[m,n]}(t;\Delta s,\epsilon)$, yielding the system of linear equations,
\begin{linenomath*} \begin{equation}
    \partial_t \,\tilde{X}_j^{\,[m]}(t) = \frac{1}{8\pi}\sum_{n=1}^Q I_{jk}^{\,[m,n]}\left(t;\Delta s,\epsilon\right) \tilde{f}_k^{\,[n]}(t), \quad m=1,\ldots, Q,
\end{equation} \end{linenomath*}
with the force and moment balance equations given in Eqs.~\eqref{eq:totalforcebalance} and \eqref{eq:totalmomentbalance} discretized via the midpoint rule as
\begin{linenomath*}
	\begin{align}
    \bm{0} & = \sum_{m=1}^Q -\Delta s \tilde{\bm{f}}^{\,[m]}(t), \\
    \bm{0} & = \sum_{m=1}^Q \Delta s \tilde{\bm{X}}^{[m]}(t)\times\left(-\tilde{\bm{f}}^{\,[m]}(t)\right) \label{eq:totalmomentbalancedisc}.
\end{align}
\end{linenomath*}
The semi-discrete form of the EIF is then
\begin{linenomath*}
	\begin{align}
\bm{0} &  = \sum_{m=1}^Q -\Delta s \tilde{\bm{f}}^{\,[m]}(t), \label{eq:forcebal2} \\
\bm{0} &  = \sum_{m=1}^Q \Delta s \tilde{\bm{X}}^{[m]}(t)\times\left(-\tilde{\bm{f}}^{\,[m]}(t)\right), \label{eq:momentbal2}\\
\frac{\tilde{\theta}^{\,[m+1]}(t)-\tilde{\theta}^{\,[m]}(t)}{\Delta s}
&    =
-\bm{e}_3\cdot\sum_{n= m}^{Q-1} \Delta s (\tilde{\bm{X}}^{[n+1]}(t)-\bm{X}^{[m+1]}(t))\times\tilde{\bm{f}}^{\,[n+1]}(t), \label{eq:elasto2} \\
\partial_t\, \tilde{X}_j^{[m]}(t) &                                                                                                                       = \frac{1}{8\pi}\sum_{n=1}^Q I_{jk}^{\,[m,n]}(t;\Delta s,\epsilon) \tilde{f}_k^{\,[n]}(t), \label{eq:fluiddyn2}\\
\mbox{where} \quad \bm{X}^{[m]}(t) &                                                                                                                    = \bm{X}^{[1]}(t)+\sum_{n=1}^{m-1} \Delta s \left[\cos\tilde{\theta}^{\,[n]}(t),\sin\tilde{\theta}^{\,[n]}(t),0\right]^T ,\nonumber \\
\quad \tilde{\bm{X}}^{[m]}(t) &                                                                                                                         = \bm{X}^{[m]}(t) +\frac{\Delta s}{2} \left[\cos\tilde{\theta}^{\,[m]}(t),\sin\tilde{\theta}^{\,[m]}(t),0\right]^T, \label{eq:semidisckinematic1} \\
\mbox{and} \quad
\partial_t \,\tilde{\bm{X}}^{[m]}(t) &                                                                                                                    = \partial_t\,\bm{X}^{[1]}(t)
+ \sum_{n                                                                                                                                               = 1}^{m-1}\Delta s\,\partial_t\,\tilde{\theta}^{\,[n]}(t)\left[-\sin\tilde{\theta}^{\,[n]}(t),\cos\tilde{\theta}^{\,[n]}(t),0\right]^T  \nonumber \\
& \quad + \frac{\Delta s}{2} \partial_t\, \tilde{\theta}^{\,[m]}(t)\left[-\sin\tilde{\theta}^{\,[m]}(t),\cos\tilde{\theta}^{\,[m]}(t),0\right]^T  \label{eq:semidisckinematic3},
\end{align}
\end{linenomath*}
where $m=1,\dots,Q-1$ in Eq.~\eqref{eq:elasto2} and $m=1,\dots,Q$ in Eq.~\eqref{eq:fluiddyn2} and \eqref{eq:semidisckinematic3}.
\sloppy

By using Eqs.~\eqref{eq:semidisckinematic1} and \eqref{eq:semidisckinematic3}, the variables $\tilde{\bm{X}}^{[1]}(t), \ldots, \tilde{\bm{X}}^{[Q+1]}(t)$ and $\bm{X}^{[2]}(t),\ldots,\bm{X}^{[Q]}(t)$
can be eliminated from Eqs.~\eqref{eq:momentbal2}, \eqref{eq:elasto2} and \eqref{eq:fluiddyn2}. The resulting system is then linear in $\partial_t\,\bm{X}^{[1]}(t),\, \partial_t\,\tilde{\theta}^{\,[1]},\ \dots,\,\partial_t\,\tilde{\theta}^{\,[Q]}, \, \tilde{\bm{f}}^{\,[1]}(t), \dots,\, \tilde{\bm{f}}^{\,[Q]}(t)$.
Hence, given the discrete configuration $\tilde{\bm{X}}^{[1]}(t),\, \tilde{\theta}^{\,[1]}(t),\dots,\,\tilde{\theta}^{\,[Q]}(t)$, the rate of change of position and angle can be found by solving a dense $(3Q+2)\times(3Q+2)$ system of linear equations. Thus, the semi-discrete system can be expressed concisely as an autonomous non-linear initial value problem,
\begin{linenomath*} \begin{equation}
    \dot{\bm{Y}}=\mathcal{F}(\bm{Y}), \quad \bm{Y}(0)=\bm{Y}_0, \label{eq:autosystem}
\end{equation} \end{linenomath*}
where $\bm{Y}(t):=[\tilde{\bm{X}}^{[1]}(t), \ \tilde{\theta}^{\,[1]}(t),\dots,\tilde{\theta}^{\,[Q]}(t) ]$ is a $ Q+2 $ column vector and $\dot{\bm{Y}}(t) = \partial_t\, \bm{Y}(t) $. By augmenting the problem with the unknown force densities, this can be written as the matrix system,
\begin{linenomath*}
	\begin{align}
    \textup{A}
    \left[ \begin{array}{c}
    	\dot{\bm{Y}}   \\
    	\tilde{\bm{f}}
    \end{array} \right]
    &= \bm{b}, \label{eq:matrixsystem-full} \\
    \text{with } \textup{A}    	&= \left[  \begin{array}{c|c}
                                                \textup{0}      & \textup{A}_E \\ \hline
                                                \textup{A}_K    & \textup{A}_H
                                            \end{array}\right], \label{eq:matrixsystem-A} \\
                    \bm{b}  	&= \left[ 0,\frac{\tilde{\theta}^{\,[2]}(t)-\tilde{\theta}^{\,[1]}(t)}{\Delta s},\dots, \frac{\tilde{\theta}^{\,[Q]}(t)-\tilde{\theta}^{\,[Q-1]}(t)}{\Delta s}, 0,0,0,\dots, 0 \right]^T, \label{eq:matrixsystem-b} \\
                    \tilde{\bm{f}}		&= \left[ \tilde{f}^{\,[1]}_1,\dots,\tilde{f}^{\,[Q]}_1, \tilde{f}^{\,[1]}_2,\dots,\tilde{f}^{\,[Q]}_2 \right]^T \label{eq:matrixsystem-c}
\end{align}
\end{linenomath*}
where $\textup{A}$ is a $(3Q+2)\times(3Q+2)$ block matrix, $\bm{b}$ is a $3Q+2$ column vector, and $ \tilde{\bm{f}} $ is a $ 2Q $ column vector, so that the concatenation $ [ \,\dot{\bm{Y}}, \tilde{\bm{f}}\, ]^T $ is a $ 3Q+2 $ column vector. The matrix blocks of $\textup{A}$ $\left(\textup{A}_E, \ \textup{A}_K, \text{ and } \textup{A}_H \right) $
encode the elastodynamic, kinematic, and hydrodynamic equations given by Eqs.~\eqref{eq:elasto2}, \eqref{eq:semidisckinematic3} and \eqref{eq:fluiddyn2} respectively. In the vector $\bm{b}$, the first entry corresponds to the moment balance on the whole filament (Eq.~\eqref{eq:momentbal2}), the subsequent $Q-1$ rows are the moment balances about each interior joint (Eq.\eqref{eq:elasto2}), and the next 2 rows correspond to the total force balance (Eq.~\eqref{eq:forcebal2}). The remaining zero entries correspond to the equivalence between the hydrodynamic and kinematic velocities (Eqs.~\eqref{eq:fluiddyn2} and \eqref{eq:semidisckinematic1}). The matrix system given in Eq.~\eqref{eq:matrixsystem-full} is solved for $ \dot{\bm{Y}} $ and $ \tilde{\bm{f}} $ at each time step using the MATLAB\textsuperscript{\textregistered} backslash command, and the resulting rates vector $ \dot{\bm{Y}} $ is integrated using the built-in variable-step, variable-order ODE solver \texttt{ode15s} \cite{shampine1997matlab}.
To demonstrate the ease of application of the EIF framework, all simulations are performed using the \MATLAB R2019a default settings for \texttt{ode15s}. In particular, the absolute and relative error tolerances are \(10^{-6}\) and \(10^{-3}\) respectively.
While this IVP exhibits some stiffness, it is less stiff than the systems produced by other methods, as the integral formulation avoid the need of additional Lagrange multipliers to ensure filament inextensibility.
At each time step, the filament is constructed according to Eq.~\eqref{eq:semidisckinematic1} with $ \Delta s = 1/Q $, ensuring that filament arclength is preserved over the course of the simulation.

\section{Single and multi-filament problems} \label{sec:single-multi-filament-problems}

In the following section, we apply the EIF framework described in Sec.~\ref{sec:model-formulation} to problems involving single or multiple filaments in the presence of body forces, surrounding flow, or undergoing self-powered propulsion.

\subsection{A single passive filament in shear flow} \label{subsec:single-fil-shear}

We investigate the dynamics of a single passive filament in a linear shear flow, ${\bm{u}^*_s(t) = \dot{\gamma}\,x_2^*(s^*,t^*) \,\bm{e}_1 }$, with shear rate $\dot{\gamma}$, where $^*$ denotes a dimensionful variable. Non-dimensionalizing with respect to the length of the filament $L$, time scale $\tau=\dot{\gamma}^{-1}$, and force density scaling $ \mu L \dot{\gamma} $ (where $\mu$ is the dynamic viscosity of the surrounding fluid), yields the non-dimensionalized equation for the hydrodynamic velocity
\begin{linenomath*} \begin{equation}
    \partial_t\,X_j(s,t) = \frac{1}{8\pi} \int_0^1 S_{jk}^{\epsilon} \left(\bm{X}(s,t),\bm{X}(s',t)\right)\, f_k(s',t)\, ds' + X_2(s,t) \,\delta_{j1}. \label{eq:hydro-shear-nondim}
\end{equation} \end{linenomath*}
Additionally, from the dimensional version of Eq.~\eqref{eq:elastEQ}, together with the scalings defined above, we obtain the non-dimensionalized elastohydrodynamic integral equation
\begin{linenomath*} \begin{equation}
    \label{eq:elasto-shear-nondim}
    \partial_s \,\theta(s,t) = -\bm{e}_3 \cdot \mathcal{V} \int_s^1 \left( \bm{X}(s',t) -\bm{X}(s,t) \right) \times \bm{f}(s',t)\, ds',
\end{equation} \end{linenomath*}
where the dimensionless viscous-elastic parameter
\begin{linenomath*} \begin{equation}
	\mathcal{V} = \frac{\mu\dot{\gamma}L^4}{EI} \label{def:V}
\end{equation} \end{linenomath*}
quantifies the ratio of viscous to elastic forces on a shear timescale, where $ EI $ is the bending rigidity of the filament. The apparent flexibility of a filament is completely characterized using $ \mathcal{V} $, with large values describing floppy fibers, and small values stiff fibers. The similarity between Eqs.~\eqref{eq:elastEQ} and \eqref{eq:elasto-shear-nondim} results in a similarly discretized form as in Eq.~\eqref{eq:elasto2}.

The hydrodynamic shear flow equation (Eq.~\eqref{eq:hydro-shear-nondim}) is semi-discretized following the steps in Sec.~\ref{sec:model-formulation} to obtain
\begin{linenomath*} \begin{equation}
    \label{eq:hydro-shear-nondim-disc}
    \partial_t\, \tilde{X}_j^{[m]}(t) = \frac{1}{8\pi} \sum_{n=1}^{Q} I_{jk}^{\,[m,n]}\left(t;\,\Delta s, \epsilon\right) \tilde{f}_k^{\,[n]}(t) + \tilde{X}^{[m]}_2(t) \,\delta_{j1}, \qquad m=1, \dots, Q.
\end{equation} \end{linenomath*}
The system of equations for a single passive filament in shear flow is thus given by Eq.~\eqref{eq:matrixsystem-full} but with the alterations
\begin{linenomath*}
	\begin{align}
	\textup{A}    	&= 	\left[  \begin{array}{c|c}
	                             \textup{0}      & \mathcal{V} \textup{A}_E \\ \hline
	                             \textup{A}_K    & \textup{A}_H
	                         	\end{array}
	                 	\right], \label{eq:matrixsystem-A-shear} \\
	\bm{b}      	&=  \left[  \vphantom{\frac{1}{1}} \right. 0,\frac{\tilde{\theta}^{\,[2]}(t)-\tilde{\theta}^{\,[1]}(t)}{\Delta s},\dots, \frac{\tilde{\theta}^{\,[Q]}(t)-\tilde{\theta}^{\,[Q-1]}(t)}{\Delta s}, 0,0,  \tilde{X}^{[1]}_2(t) ,\dots, \tilde{X}_2^{[Q]}(t) ,0,\dots,0  \left. \vphantom{\frac{1}{1}} \right]^T, \label{eq:matrixsystem-b-shear}
\end{align}
\end{linenomath*}
and where the vector of unknowns $[ \,\dot{\bm{Y}}, \tilde{\bm{f}}\, ]^T$ remains unchanged from Eq.~\eqref{eq:matrixsystem-full}.

\subsection{A single passive filament sedimenting under gravity} \label{subsec:single-fil-sedimenting}

The EIF method also allows for implementation of a body force to the system. In this section, the simulation of passive filaments sedimenting under gravity is considered. Assuming uniform mass per unit length $ \rho $, the force per unit length due to gravity acting upon the filament is $ -\rho g \bm{e}_2 $. Non-dimensionalizing with respect to the filament length $ L $, time scaling $ \tau = \mu L^4/EI $, and force density $ \mu L/\tau $, addition of the gravitational force density produces the force and moment balance equations
\begin{linenomath*}
	\begin{align}
	\int_0^1 \left(-\bm{f}(s',t) -\mathcal{G}\bm{e}_2\right) \,ds' &= \bm{0} \label{eq:forcebal-grav}\\
	\int_0^1 \left( \bm{X}(s',t)-\bm{X}_c(t) \right) \times \left(-\bm{f}(s',t)-\mathcal{G}\bm{e}_2\right)\,ds' &= \bm{0},	 \label{eq:mombal-grav}
\end{align}
\end{linenomath*}
where $ \bm{X}_c(t) $ is the center of gravity of the filament, $ \bm{f}(s,t) $ is the force per unit length the filament exerts on the fluid, and $ \mathcal{G} $ is the elasto-gravitational parameter
\begin{linenomath*} \begin{equation}
	\mathcal{G} = \frac{\rho g L^3}{EI}. \label{def:calG}
\end{equation} \end{linenomath*}
Following a derivation similar to that presented in Sec.~\ref{sec:model-formulation}, the elastohydrodynamic equation is found as
\begin{linenomath*} \begin{equation}
	\partial_s\,\theta(s,t)= -\bm{e}_3 \cdot \int_{s}^{1} \left( \bm{X}(s',t)-\bm{X}(s,t) \right)\times \left( \bm{f}(s',t)+\mathcal{G}\bm{e}_2 \right)\, ds', \label{eq:elasto-grav}
\end{equation} \end{linenomath*}
which, along with Eqs.~\eqref{eq:forcebal-grav} and \eqref{eq:mombal-grav}, has spatially discretized form
\begin{linenomath*}
\begin{align}
	\bm{0} &= \sum_{m=1}^{Q} \Delta s \left(-\tilde{\bm{f}}^{\,[m]}(t)-\mathcal{G}\bm{e}_2\right), \label{eq:fbal-grav-disc}\\
	\bm{0} &= \sum_{m=1}^{Q} \Delta s \,\left( \tilde{\bm{X}}^{[m]}(t) - \bm{X}_c(t) \right)\times \left(-\tilde{\bm{f}}^{\,[m]}(t)-\mathcal{G}\bm{e}_2 \right), \label{eq:mbal-grav-disc} \\
  \frac{\tilde{\theta}^{\,[m+1]}(t)-\tilde{\theta}^{\,[m]}(t)}{\Delta s} &= -\bm{e}_3\cdot \sum_{n=m}^{Q-1} \Delta s \left( \tilde{\bm{X}}^{[n+1]}(t)-\bm{X}^{[m+1]}(t) \right)\times\left( \tilde{\bm{f}}^{\,[m+1]}(t)+\mathcal{G}\bm{e}_2 \right). \label{eq:elasto-grav-disc}
\end{align}
\end{linenomath*}
As in Sec.~\ref{subsec:semi-disc-of-eif}, we form a matrix system encoding Eqs.~\eqref{eq:fbal-grav-disc}--\eqref{eq:elasto-grav-disc} along with Eqs.~\eqref{eq:fluiddyn2} and \eqref{eq:semidisckinematic3}
\begin{linenomath*} \begin{equation}
	\textup{A}_g\,
    \left[ \begin{array}{c}
		\dot{\bm{Y}}   \\
		\tilde{\bm{f}}
	\end{array} \right]
	 = \bm{b}_g, \label{eq:matrixsystem-full-grav}
\end{equation} \end{linenomath*}
which can be solved to find the filament velocities $ \partial_t\,\bm{X}_0(t), \, \partial_t\,\tilde{\theta}^{\,[m]}(t) $ and force densities $ \tilde{\bm{f}}^{\,[m]}(t) $ for $ m=1,\dots,Q $. The matrix $ \textup{A}_g $ has the same block form as $ \textup{A} $, but with an alteration in the first row of the elasticity block  $ \textup{A}_E $ due to do the inclusion of the center of gravity in the total moment balance equation (Eq.~\ref{eq:mbal-grav-disc}). The right hand side vector is constructed as $ \bm{b}_g = \bm{b} + \hat{\bm{b}} $, where $ \hat{\bm{b}} $ encodes all the changes required to the right hand side vector $ \bm{b} $ (given in Eq.~\eqref{eq:matrixsystem-b}) after expansion and rearrangement of Eqs.~\eqref{eq:fbal-grav-disc}, \eqref{eq:mbal-grav-disc}, and \eqref{eq:elasto-grav-disc}.

\subsection{A single active filament} \label{subsec:single-fil-swimming}

For the problems considered in Secs.~\ref{subsec:single-fil-shear} and \ref{subsec:single-fil-sedimenting}, we can consider active filaments by including a time-dependent moment density term added to the elastodynamic formulation given in Eq.~\eqref{eq:contelasto}. The moment balance in Eq.~\eqref{eq:momentdensity} is extended
\begin{linenomath*} \begin{equation}
	\partial_s\,M(s,t) + \bm{e}_3\cdot \left(\partial_{s}\,\bm{X}(s,t) \times \bm{N}(s,t)\right) + \mathfrak{m}(s,t)=0, \label{eq:active-moment-density}
\end{equation} \end{linenomath*}
where $\mathfrak{m}(s,t)$ is the moment per unit length that drives the actuation of the filament. Continuing the non-dimensionalization as in Sec.~\ref{sec:model-formulation}, with time scaling $ \tau = \omega^{-1} $, yields the non-dimensionalized elastohydrodynamic equation for a single active filament
\begin{linenomath*} \begin{equation}
	\partial_s\,\theta(s,t) - \mathcal{S}^{\,4} \int_s^1 \mathfrak{m}(s',t)\,ds' = -\bm{e}_3\cdot\, \mathcal{S}^{\,4} \int_s^1 \left( \bm{X}(s',t) -\bm{X}(s,t) \right)\times \bm{f}(s',t)\,ds', \label{eq:cont-swimming-mom}
\end{equation} \end{linenomath*}
where $ \mathcal{S} $ is the dimensionless swimming number, defined
\begin{linenomath*} \begin{equation}
	\mathcal{S} = L\left( \frac{\mu \omega}{EI} \right)^{1/4},
\end{equation} \end{linenomath*}
with $ \omega $ being the angular velocity of the swimming beat prescribed to the filament. Note that the swimming parameter is different from the commonly-used sperm number $ \textup{Sp} $, which has dependence on a chosen resistance coefficient. Following work by Gad\^elha \textit{et al.} \cite{gadelha2010nonlinear} and Montenegro-Johnson \textit{et al.} \cite{montenegro2015spermatozoa}, the active moment $ \mathfrak{m}(s,t) $ is set as a traveling wave with amplitude $ \mathfrak{m}_0 $, wave number $ k $ and angular frequency $ \omega $. The semi-discretized form of Eq.~\eqref{eq:cont-swimming-mom} is
\begin{linenomath*}
	\begin{align}
	\frac{\tilde{\theta}^{\,[m+1]}(t)-\tilde{\theta}^{\,[m]}(t)}{\Delta s} - \mathcal{S}^{\,4}\, \int_{\tilde{s}^{\,[m]}}^{1} \mathfrak{m}(s',t)\,ds'& \nonumber \\
	&\hspace{-2cm}= -\bm{e}_3\cdot \mathcal{S}^{\,4} \sum_{n=m}^{Q-1} \Delta s \left( \tilde{\bm{X}}^{[n+1]}(t)-\bm{X}^{[m+1]}(t) \right) \times\tilde{\bm{f}}^{\,[n+1]}(t), \label{eq:swimming-moment-disc}
\end{align}
\end{linenomath*}
for $ m=1,\dots,Q $. In the following section, we consider the how the presented framework can be extended to model systems of multiple filaments.

\subsection{Systems of multiple filaments} \label{subsec:multi-fil-stationary-fluid}

The EIF can be used to simulate the dynamics of large groups of filaments, accounting for the non-local hydrodynamic interactions between them. For each of the problems presented in Secs.~\ref{subsec:single-fil-shear}, \ref{subsec:single-fil-sedimenting}, and \ref{subsec:single-fil-swimming}, the kinematic and elastodynamic equations apply to each filament in the system individually. The hydrodynamic equations are extended so that interactions between all filaments are considered. For a system of $ N $ passively relaxing filaments, this equation reads
\begin{linenomath*} \begin{equation}
	\partial_t\,X_j^{\{\alpha\}}(s,t) = \frac{1}{8\pi} \sum_{\beta=1}^{N} \int_{0}^{1} S_{jk}^{\varepsilon} \big( \bm{X}^{\{\alpha\}}(s,t), \bm{X}^{\{\beta\}}(s',t) \big)\,f_k^{\,\{\beta\}}(s',t)\,ds', \qquad \beta = 1,\dots, N, \label{eq:multi-fil-hydrodynamics}
\end{equation} \end{linenomath*}
where the superscript $ \{\alpha\} $ in this continuous equation refers to the $ \{\alpha\} $\textsuperscript{th} filament in the system.
Whilst three-dimensional hydrodynamic effects are computed, the kinematics and hence elasticity of the multiple filament problem remains two-dimensional, ensuring planarity.
Modification of Eq.~\eqref{eq:multi-fil-hydrodynamics} to consider external flows or body forces follows from the single filament derivations presented in Secs.~\ref{subsec:single-fil-shear}, \ref{subsec:single-fil-sedimenting}, and \ref{subsec:single-fil-swimming}, and non-dimensionalizing yields the same governing dimensionless parameters $ \mathcal{V} $, $ \mathcal{G} $, and $ \mathcal{S} $ respectively. Discretization of the resulting equations is performed in the same manner, producing an $ {N(3Q+2)\times N(3Q+2)} $ linear system similar in structure to those presented in each of the single filament problem descriptions. Full details of the equations as well as the associated linear systems for each of the multiple-filament problems are given in the Supplemental Material Sec.~SII.

\section{Model verification}\label{sec:verification}

To verify the accuracy and efficacy of the EIF, we compare the computed dynamics of a single relaxing filament between the proposed method and a high resolution bead and link method (BLM) formulation. Based upon the work of Jayaraman \textit{et al.} \cite{jayaraman2012autonomous}, the BLM accounts for non-local hydrodynamic interactions, and when highly resolved, provides accurate solutions (details provided in the Supplemental Material Sec.~SIII). For this reason, the BLM is used to both verify the proposed EIF method, and as a reliable point of comparison between other existing methods. We consider the case of a filament, bent into a parabola, with initial condition $ \bm{Y} $ constructed by sampling symmetrically from the curve $ y=0.5x^2 $, and ensuring unit arclength. This filament is then allowed to relax with no external forcing. Motion of the filament in this scenario is due to the constitutive bending moments along the arclength, given in Eq.~\eqref{eq:elasto2}. This experiment is simulated using each of four methods:
\begin{enumerate}
    \item EIF-RSM: the proposed EIF method, which uses regularized stokeslets to account for \textit{non-local} hydrodynamic interactions,
    \item EIF-RFT: the EIF method, with resistive force theory in place of the method of regularized stokeslets to model \textit{local} hydrodynamic interactions (refer to the associated Supplemental Material Sec.~SIV for full details),
    \item MGG: the original angle formulation method by Moreau, Giraldi and Gad\^elha \cite{moreau2018asymptotic}, which uses resistive force theory to model \textit{local} hydrodynamic interactions,
    \item BLM: the bead and link method, which accounts for \textit{non-local} hydrodynamic interactions.
\end{enumerate}
We include the EIF-RFT method to verify the equivalence of the present implementation with that of Moreau \textit{et al.} \cite{moreau2018asymptotic} under the reduction to local hydrodynamics. In all simulations in this paper the regularization parameter for use in the method of regularized stokeslets is chosen as $ \varepsilon= 0.01 $.

\begin{figure}
	\centering
	\subfloat{\includegraphics[clip,trim=0.3cm 0.2cm 0.5cm 0cm,width=0.95\textwidth]{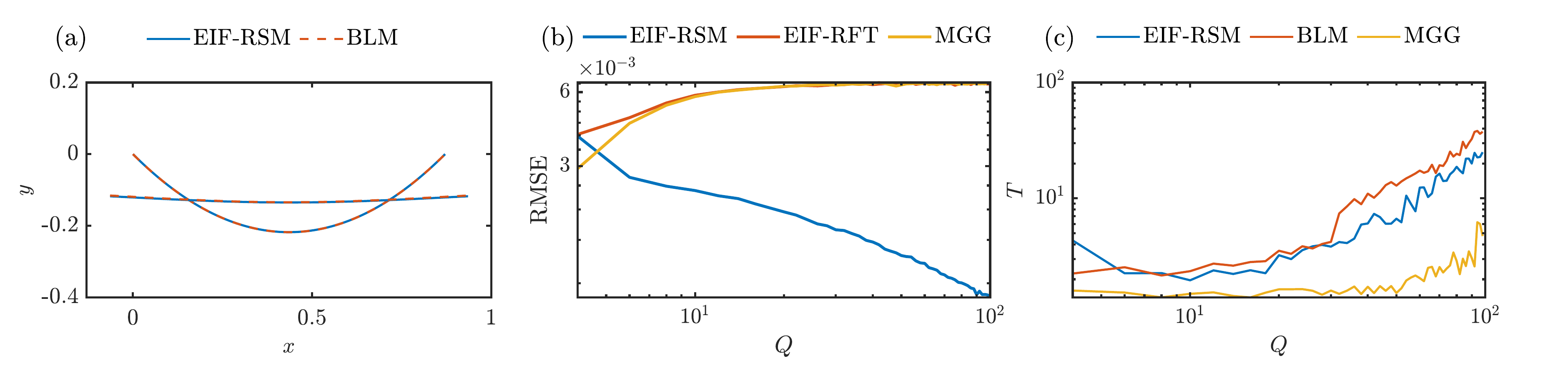} \label{figure2a}}
	\subfloat{ \label{figure2b}}
	\subfloat{ \label{figure2c}}
	\caption{
     	Error convergence of three EIF methods against a very high resolution ($Q=200$) bead and link model (detailed in Sec.~SIII of the associated Supplemental Material). In (a), the geometry of the relaxing rod experiment is displayed, comparing the shape at $t=0$ and $t=0.02$ for the EIF-RSM with $Q=100$ to the high-resolution BLM with $Q=200$. In (b), the root mean squared error (RMSE) is calculated between the Cartesian solution data for a relaxing filament at $t=0.02$ modeled using a finely-discretized BLM formulation and (1) EIF-RSM, (2) EIF-RFT i.e. the EIF method but using a resistive force approximation (see Sec.~SIV of the Supplemental Material), and (3) MGG, the EIF method proposed by Moreau \textit{et al.} \cite{moreau2018asymptotic}. (c) Total wall time $T$ against level of discretization $Q$ for the EIF-RSM, MGG and BLM methods.
	}
	\label{figure2}
\end{figure}

The geometric configuration of the relaxing filament is simulated between $ t=0 $ and $t=0.02$, using both a high-resolution ($ Q=200 $) BLM formulation and the EIF-RSM method with $ Q=100 $. The initial and final filament shapes from each simulation are displayed in Fig.~\ref{figure2a}, which show excellent agreement between these formulations. Quantitative comparison is shown in Fig.~\ref{figure2b}, where we plot the root mean squared error (RMSE) between the position of the filament described by each model against that of the high-resolution BLM. Here excellent convergence is evident, with the RMSE being minimal for even small values of $Q$. Comparisons with MGG and EIF-RFT are also presented in this figure. While it might be counter-intuitive that the local hydrodynamic models initially increase in error with $Q$, this is due to drift in the center of mass that the non-local methods correctly capture. Increasing $Q$ in the local methods leads to increasingly resolved shape of the relaxing filament, leading to the convergence of this error. These comparisons highlight the change in dynamics when considering the inclusion of non-local hydrodynamics for even a simple single filament problem.

Moreau \textit{et al.} demonstrated the significant reduction in computational cost achieved when formulating elastohydrodynamic problems as an integro-differential equation. With the added complexity of including non-local hydrodynamic modeling, the EIF-RSM still performs very well. In Fig.~\ref{figure2c} we plot logarithmic comparisons of the simulation runtime for each of EIF-RSM, MGG, and BLM formulations. The walltime recorded is the total computational time for the method including setup time. For the tangent angle formulation methods, the majority of this time is accounted for by the linear solve at each time step.

It it unsurprising that the \textit{local} hydrodynamic formulation (MGG) outperforms the methods with \textit{non-local} interactions due to reduced complexity of the problem. The EIF-RSM and BLM method perform on par for increasing $Q$. However, the BLM method requires large numbers of beads in order to accurately capture the correct filament dynamics (see Sec.~SIII of the provided Supplemental Material), whereas the EIF-RSM can achieve similar accuracy with fewer than half of the degrees of freedom required by the BLM (see Fig.~\ref{figure2b}).

\section{Results} \label{sec:results}

In the following section, results from the numerical experiments outlined in Sec.~\ref{sec:single-multi-filament-problems} are presented. We begin by examining the dynamics of single passive filaments in shear flow and sedimenting under gravity, and active filaments with prescribed internal moments. Additionally, we investigate larger arrays of passive filaments, again in shear flow and undergoing sedimentation, and systems of multiple swimming filaments. The regularization parameter is $ \epsilon=0.01 $ and, unless otherwise stated, $ Q=40 $, with segment lengths $ \Delta s = 1/Q $. This choice of $ Q $ is motivated by the convergence results given in the Supplemental Material Sec.~SI. Simulations are run on a computer equipped with an Intel i7-8750H processor and 16GB of RAM.

\subsection{Results: a single passive filament in shear flow}\label{subsec:results-single-fil-shear}

It is well known that for critical values of the characteristic parameter, a filament in shear flow exhibits shape buckling \cite{liu2018morphological,tornberg2004simulating} due to a stress difference across the fiber while under compression by the fluid. Dynamics of a filament in shear flow are simulated by solving Eq.~\eqref{eq:matrixsystem-full} with A and $ \bm{b} $ given by Eqs.~\eqref{eq:matrixsystem-A-shear} and \eqref{eq:matrixsystem-b-shear}, and characterized by parameter $ \mathcal{V} $ (Eq.~\eqref{def:V}).

We initialize our filament as a straight rod with a small perturbation following the method of Young \cite{young2009hydrodynamic}, writing the initial angular configuration as
\begin{linenomath*}
\begin{equation}
	\tilde{\theta}^{[n]}(0) = \theta_0 + \Delta \theta_0 \left( \frac{\left(\tilde{s}^{[n]}\right)^3}{3} - \frac{\left(\tilde{s}^{[n]}\right)^4}{2} + \frac{\left(\tilde{s}^{[n]}\right)^5}{5} \right), \label{eq:shear-int}
\end{equation}
\end{linenomath*}
for initial angle $ \theta_0 $ and small perturbation parameter $ \Delta \theta_0 $.  As discussed by Tornberg \& Shelley \cite{tornberg2004simulating}, prescribing a small perturbation to a straight filament can drastically change the dynamics from rotational Jeffery orbits to interesting buckling phenomena.

\begin{figure}[tp]
	\centering
	\subfloat{ \includegraphics[width=0.98\textwidth]{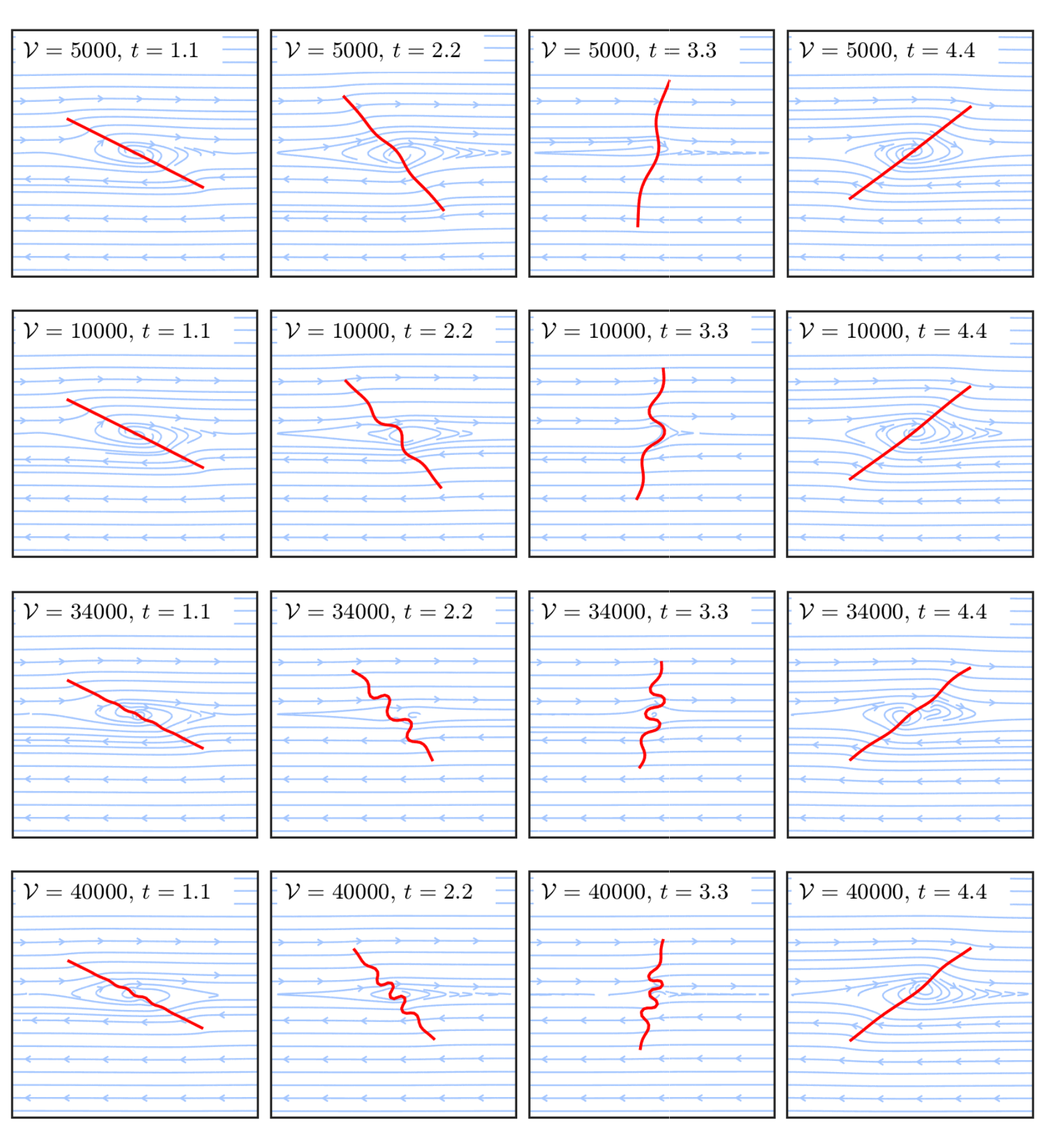} }
    \vspace{0.2cm}
    \caption{
    	Dynamics of a single filament in shear flow. Each row displays the geometric configuration of the filament, characterized by different values of $\mathcal{V}$, as it rotates in a shear flow over time. Streamlines indicate the direction of the surrounding fluid flow. In each case, the filament is modeled using $Q=40$ segments, with initial shape parameters $ \theta_0=0.9\pi $ and $ \Delta\theta_0=0.1 $ (Eq.~\eqref{eq:shear-int}). For a given initial condition, the size of $\mathcal{V}$ completely determines the level of buckling the filament exhibits as it is compressed during the rotation. For large values, high-order buckling modes appear, as in the third and fourth rows of the above diagram.
    }
	\label{figure3}
\end{figure}

In Fig.~\ref{figure3}, we demonstrate how changing the value of $ \mathcal{V} $ affects the amount of buckling a filament experiences. The fiber is perturbed with $ \theta_0=0.9\pi $ and $ \Delta \theta_0=0.1 $. As the filament rotates, buckling modes form, which are directly linked to the size of $ \mathcal{V} $ and the choice of perturbation (Eq.~\eqref{eq:shear-int}).
Large values produce high-order buckling modes, which can be seen in the fourth row of Fig.~\ref{figure3}. Conversely, comparatively small values produce no buckling and the filament rotates through a standard Jeffery orbit. For $\mathcal{V} = 5\times 10^3$, a first-order buckling mode begins to form, visible in the first row of Fig.~\ref{figure3}. The amplitude of the buckling increases as $ \mathcal{V} $ increases until higher-order modes are induced. Tornberg \& Shelley \cite{tornberg2004simulating} examined buckling governed by an effective viscosity parameter $ \bar{\mu} = 8\pi\mathcal{V} $, producing filament shapes akin to the second row of Fig.~\ref{figure3}.

The buckling problem of a flexible filament in shear flow has multiple solutions \cite{becker2001instability,liu2018morphological}. As noted by Tornberg \& Shelley \cite{tornberg2004simulating}, the choice of initial perturbation to the filament shape can preferentially lead to one of the solutions.
For example, the particular initial condition considered in \cite{tornberg2004simulating} produces a reflected filament shape when changing the sign of the perturbation.
Additionally, for the high-order buckling modes, increasing values of \(Q\) must be chosen in order to resolve the filament dynamics. However, given an initial condition that does not uniquely determine a specific solution branch (as for this problem with large values of \(\mathcal{V}\)), the buckled solution becomes sensitive to the choice of discretization. Details of the convergence of the method for this problem is contained within Sec.~SI.1 of the Supplemental Material.

\begin{figure}[tp]
	\centering
	\subfloat{\includegraphics[width=0.9\textwidth]{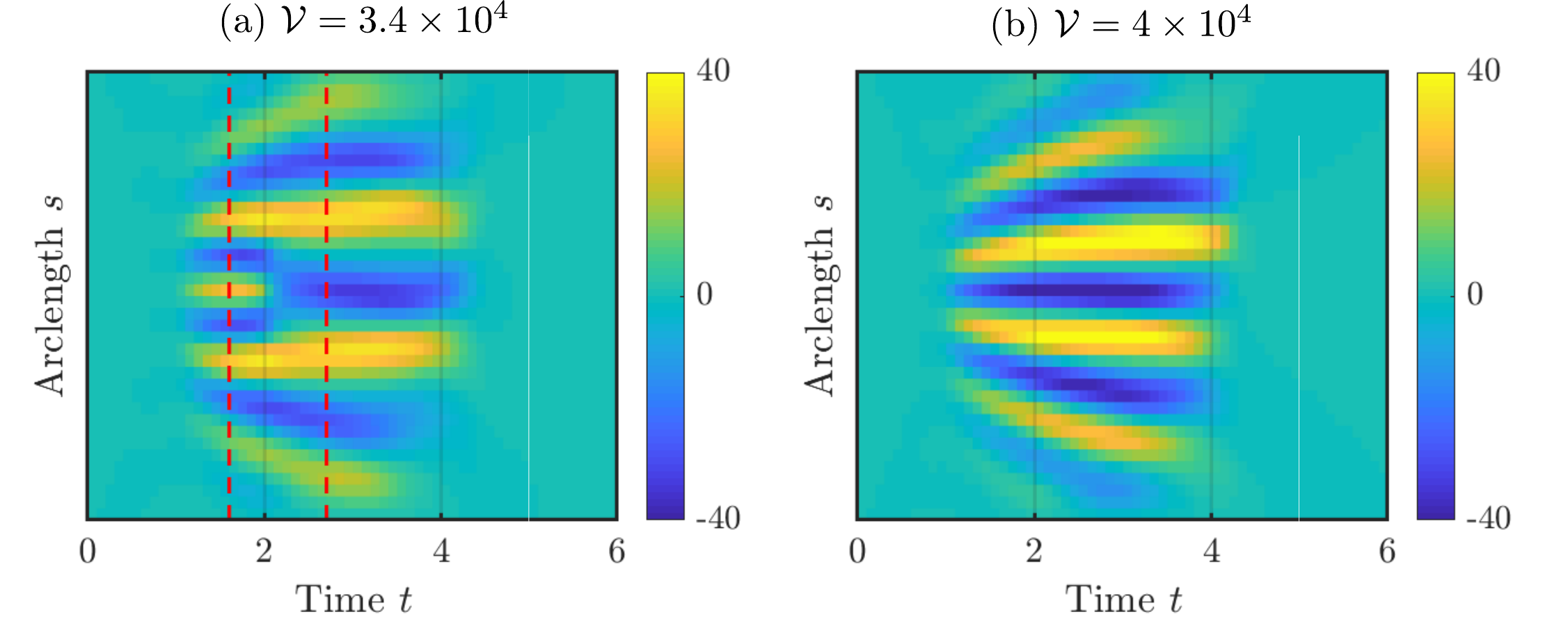} \label{figure4a} }
	\subfloat{\label{figure4b}}
	\caption{
    Time evolution of curvature for two choices of the dimensionless group \(\mathcal{V}\). (a) \(\mathcal{V}=3.4\times 10^4\), (b) \(\mathcal{V}=4\times 10^4\). Note the collapse of a higher order mode to a lower order mode in the case of \(\mathcal{V}=3.4\times 10^4\) (panel (a), with the red dashed lines indicating the zone of collapse).
	}
	\label{figure4}
\end{figure}

At critical values of $\mathcal{V}$, unstable buckling can occur. For example, when ${\mathcal{V}=3.4\times 10^4}$, a higher-order buckling mode initially forms (row three of Fig.~\ref{figure3}) but collapses into a lower-order mode. In Fig.~\ref{figure4}, the filament evolution highlighted in Fig.~\ref{figure3} is displayed as a curvature plot in arclength and time. The unstable higher-order mode can be clearly seen collapsing into a lower-order configuration, indicated by the two dotted lines in Fig.~\ref{figure4a}. Increasing $ \mathcal{V} $ further ensures that a higher-order mode is stable throughout the compression phase (Fig.~\ref{figure4b}).

\begin{figure}[!]
	\centering
	\subfloat{\includegraphics[width=0.99\textwidth]{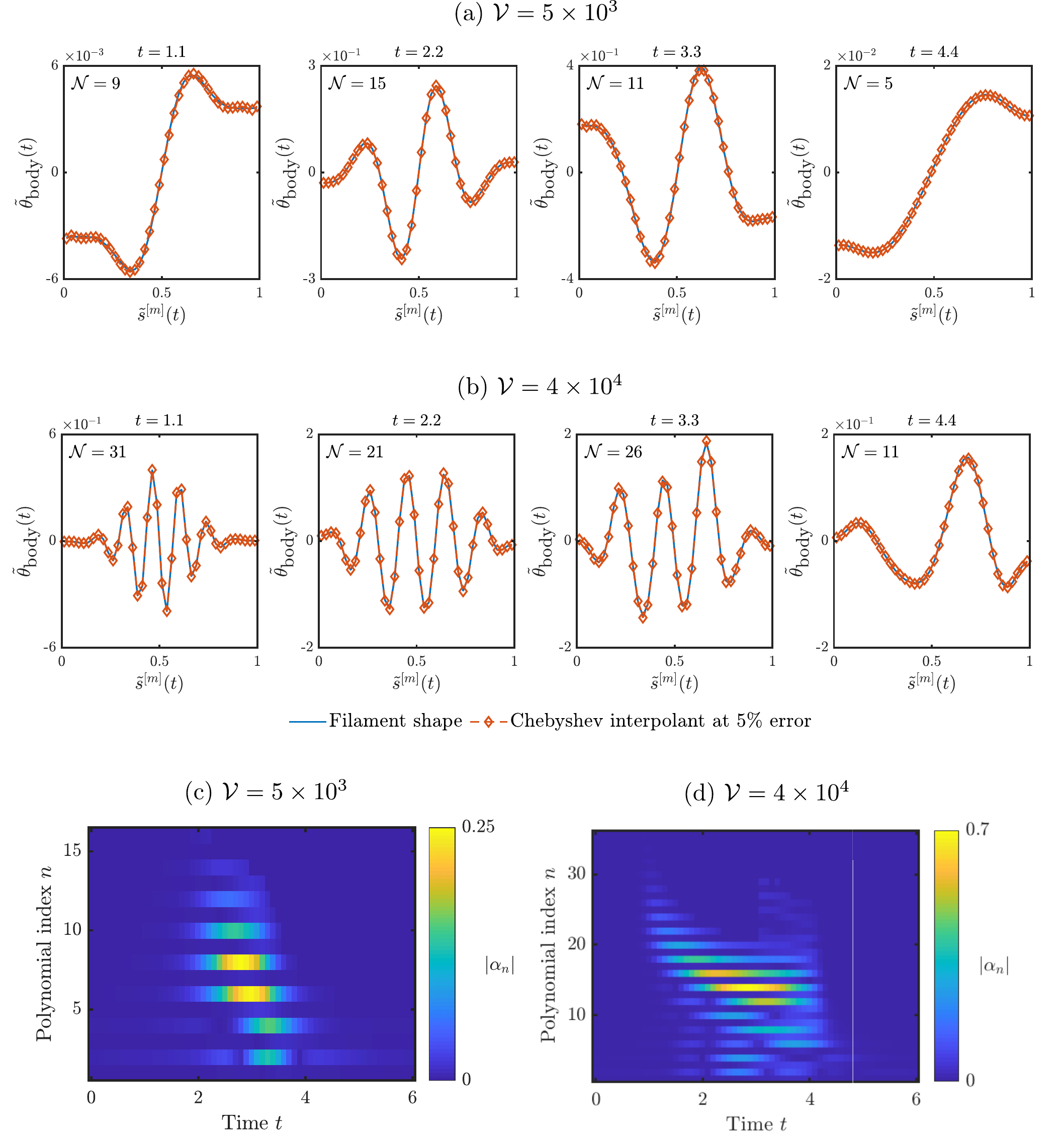} \label{figure6a}} \\
	\subfloat{\label{figure6b} } \\
	\subfloat{\label{figure6c}}
	\subfloat{\label{figure6d}} \\ \vspace{-1cm}
 	\caption{
		Analysis of the buckling modes of a filament in shear flow via polynomial interpolation. At select time points during the filament rotation, we fit a series of \(\mathcal{N}\) Chebyshev polynomials \(\mathcal{T}_n(\tilde{\theta}_\textup{body})\) (\(n=1,\hdots,\mathcal{N}\)), with \(\mathcal{N}\) chosen so that the tangent angle curve is interpolated within a 5\% error bound. The polynomial coefficients are calculated using the \texttt{chebfun} package for \MATLAB. In (a) and (b), the tangent angle of a single filament undergoing buckling is captured relative to the body frame at four time points during its rotation. Each subplot indicates the shape of the filament via $ {\tilde{\theta}_{\text{body}} = \tilde{\theta}^{[m]}(t) - \overline{\theta}(t)}$ , the values of the interpolating Chebyshev polynomial at discrete segment midpoints $\tilde{s}^{[m]}$, for each degree \(n\). In (c) and (d), the number and magnitude of Chebyshev coefficients for each of the polynomials is presented, for $\mathcal{V}=5\times 10^3$ and $4\times 10^4$ respectively.
	}
	\label{figure6}
\end{figure}

The perturbation to the filament shape induced by buckling can be investigated by considering the evolving body-frame tangent angles $ {\tilde{\theta}^{\,[m]}_{\text{body}} = \tilde{\theta}^{[m]}(t) - \overline{\theta}(t)}$, where
\begin{linenomath*}
\begin{equation}
	\overline{\theta}(t) = \frac{1}{Q}\sum_{m=1}^{Q}\tilde{\theta}^{\,[m]}(t) \label{eq:av-theta}
\end{equation}
\end{linenomath*}
is the mean filament angle. By fitting Chebyshev polynomials to $ \tilde{\theta}^{\,[m]}_{\text{body}}  $ along the filament at a time $t$, allowing for $5\%$ interpolation error, we can assess the evolution of both the order of Chebyshev polynomials required and their associated magnitude. In Figs.~\ref{figure6a} and \ref{figure6b}, we show the results of this fitting process for two choices of $\mathcal{V}$. An increase in $ \mathcal{V} $ requires a commensurate increase in polynomial order required, illustrated by examining the Chebyshev coefficients in Figs.~\ref{figure6c} and \ref{figure6d}.

\subsection{Results: a single passive filament sedimenting under gravity} \label{subsec:results-single-fil-sedimenting}

\begin{figure}[t]
    \centering
    \includegraphics[width=\textwidth]{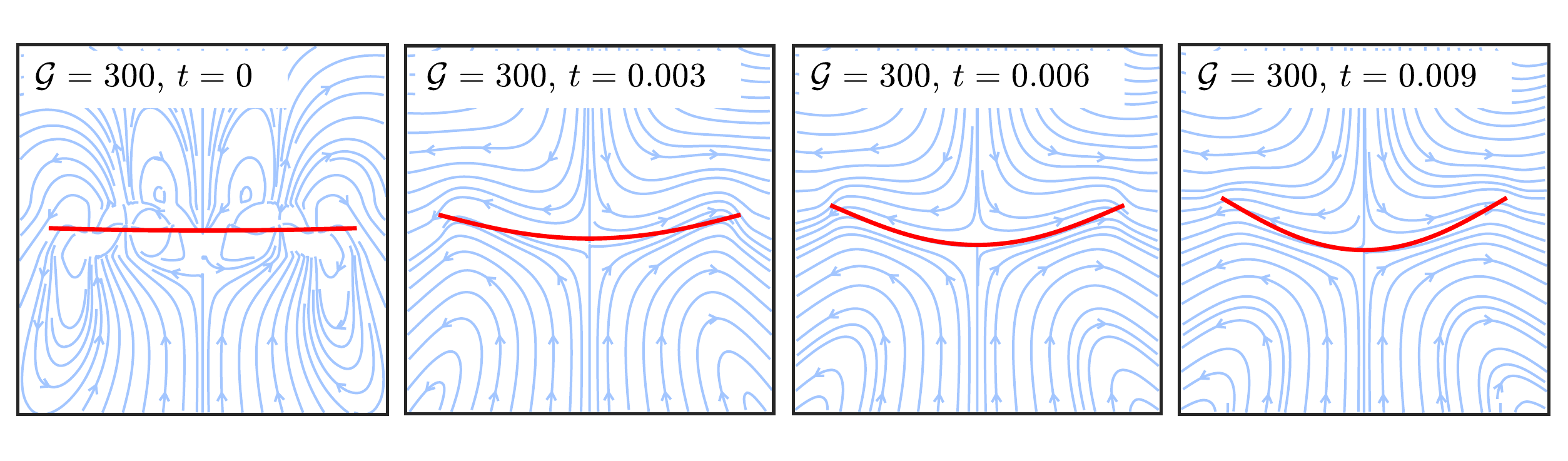}
    \caption{
    	Dynamics of an elastic filament sedimenting under gravity. Axes are centered on the center of mass of the filament at the corresponding time point.
    }
    \label{figure7}
\end{figure}

Following Sec.~\ref{subsec:single-fil-sedimenting}, simulations of a single passive filament sedimenting under gravity (with no background flow) are considered. Filament dynamics in this case are simulated by solving Eq.~\eqref{eq:matrixsystem-full-grav} and characterized by the dimensionless elasto-gravitational parameter $ \mathcal{G} $ (Eq.~\eqref{def:calG}).

By sampling \(\theta\) from the very-low amplitude parabola $ {y=1\times 10^{-7}x^2 }$, the filament geometry is initialized with unit arclength, and pre-solved with a coarse discretization ($ Q=10 $) until the shape is sufficiently curved so that a higher-resolution representation can be employed (full details are provided in the Supplemental Material Sec.~SI.2). In the following results, \(Q=40\) is used for the upscaled initial condition.

\begin{figure}[t]
    \centering
    \includegraphics[width=\textwidth]{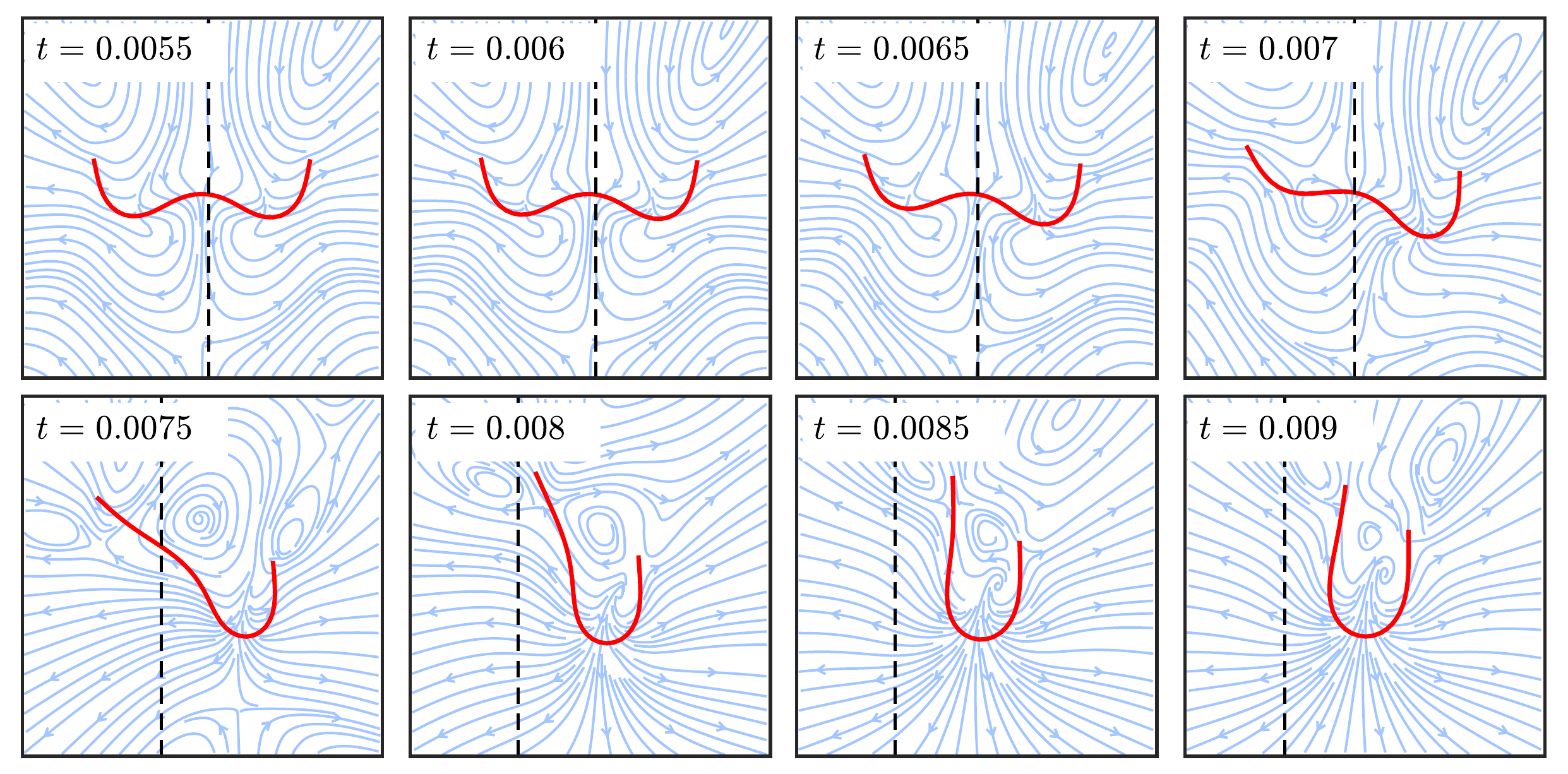}
    \caption{
    	Dynamics of a very flexible filament sedimenting under gravity. Here, for $\mathcal{G}=3500$, the filament first assumes a meta-stable ``W'' profile, before instability along the arclength transforms causes transition to the stable ``U'' configuration. In each panel, axes are centered on the center of gravity of the filament at the corresponding time point. The dashed line indicates the position of the center of gravity at $t=0$.
    }
    \label{figure8}
\end{figure}

For different choices of $\mathcal{G}$, various sedimentary buckling modes can be observed. In Fig.~\ref{figure7}, a stable ``U'' filament configuration emerges over time. For large values of $\mathcal{G} > 3000$, a metastable ``W'' configuration forms, which transitions into a stable ``U'' horseshoe shape (see Fig.~\ref{figure8}, in which $ \mathcal{G}=3500 $). Such behavior has previously been observed by Delmotte \textit{et al.} \cite{delmotte2015general} and Cosentino Lagomarsino \textit{et al.} \cite{consentino2005hydrodynamic}, who witnessed buckling at the same threshold value for their identical elasto-gravitational parameter. This transition shifts the filament's center of gravity to the left, creating an asymmetry which is then partially resolved as the horseshoe equilibrium configuration is reached.

\subsection{Results: a single active filament}\label{subsec:results-single-active-fil}

In the following section, we consider swimming in a stationary fluid caused by two types of traveling-wave moment densities, (1) sperm-like sinusoidal motion $ \mathfrak{m}_1(s,t) $, and (2) a worm-like beating pattern $ \mathfrak{m}_2(s,t) $, given respectively by
\begin{linenomath*}
	\begin{align}
	\mathfrak{m}_1(s,t) &= \mathfrak{m}_0 \, s \, \cos(ks-t), \label{eq:sperm-beat} \\
	\mathfrak{m}_2(s,t) &= \mathfrak{m}_0\, \cos(ks-t), \label{eq:worm-beat}
\end{align}
\end{linenomath*}
where $ \mathfrak{m}_0 $ and $ k $ are the dimensionless amplitude and wave number respectively. These swimmers are initialized by sampling \(\theta\) from a low-amplitude parabola $ y=1\times 10^{-3}x^2 $ and constructed as before, ensuring unit arclength.

For appropriately small choices of \(\mathfrak{m}_0\), substitution of Eq.~\eqref{eq:sperm-beat} or  \eqref{eq:worm-beat} into Eq.~\eqref{eq:swimming-moment-disc} can induce swimming in a filament in stationary surrounding flow.
Fixing the wave number $k$, the relationship between filament elasticity and choice of driving force (governed by $ \mathfrak{m}_0 $) is investigated. The Velocity Along a Line (VAL) is a measure of the swimming speed of a filament for  a chosen $ \mathcal{S} $ and $ \mathfrak{m}_0 $ pair, calculated via
\begin{linenomath*} \begin{equation}
	\textup{VAL} = \frac{\| \bm{X}_0^{(j)} - \bm{X}_0^{(j-1)} \|}{T}, \label{def:VAL}
\end{equation} \end{linenomath*}
in which $ T = 2\pi $ is the period of the driving wave and $ \bm{X}_0^{(j)} $ represents the position of the leading point of the filament after it has traveled $ j $ wavelengths, with $ j $ chosen such that the filament has established a regular motion after beginning to swim.

\begin{figure}
	\centering
	\includegraphics[width=0.75\linewidth]{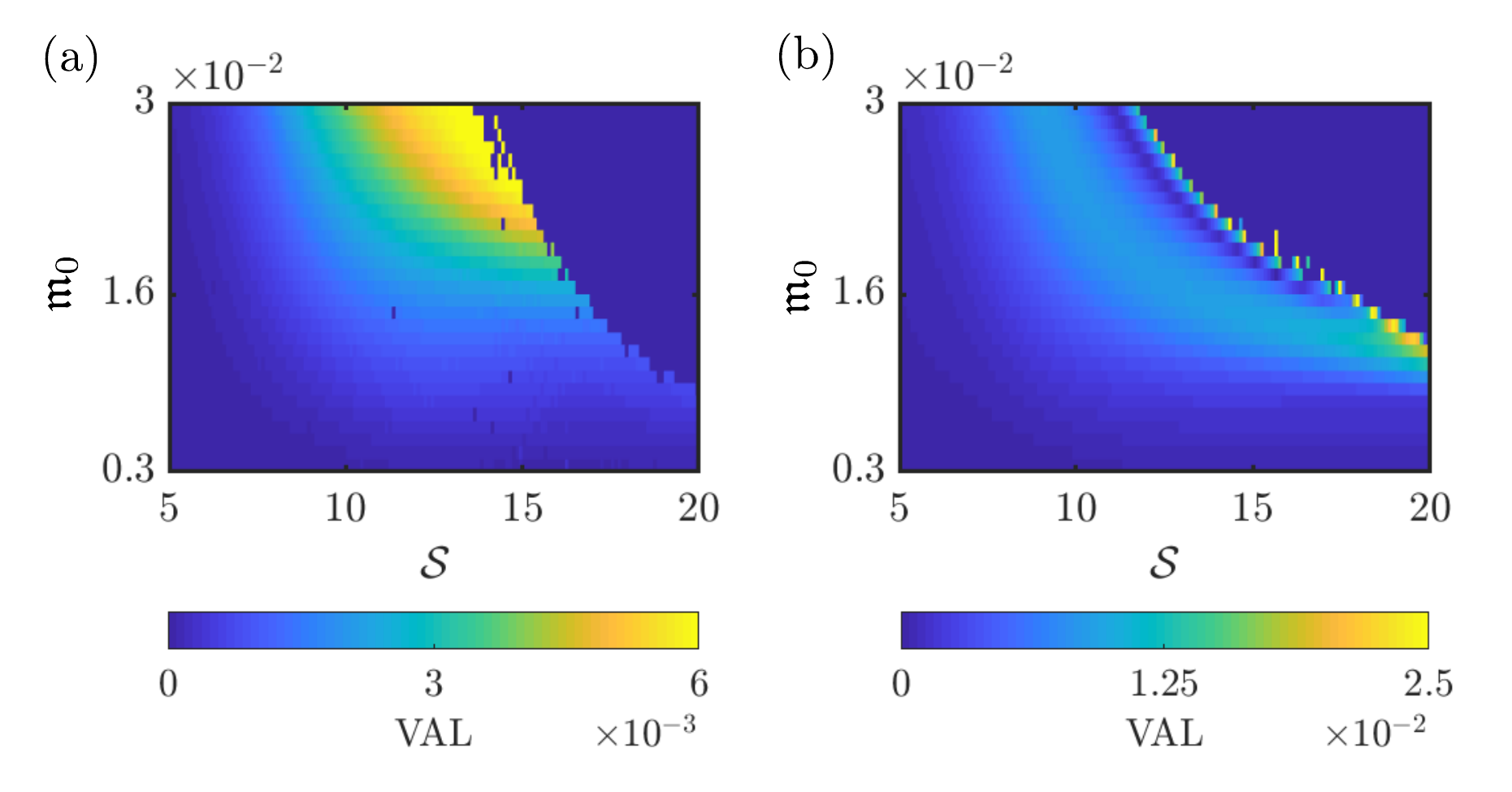}
	\caption{
		Swimming speed of a filament and its relation to parameter choices. The Velocity Along a Line (VAL) is calculated for various pairs of filament swimming number $ \mathcal{S} $ and traveling-wave amplitude $ \mathfrak{m}_0 $. Here, VAL is shown for (a) a sperm-like swimmer (Eq.~\eqref{eq:sperm-beat}) and (b) a worm-like swimmer (Eq.~\eqref{eq:worm-beat}), each with fixed wave number $ k=4\pi $.
	}
	\label{figure9}
\end{figure}

\begin{figure}
	\centering
	\includegraphics[width=\textwidth]{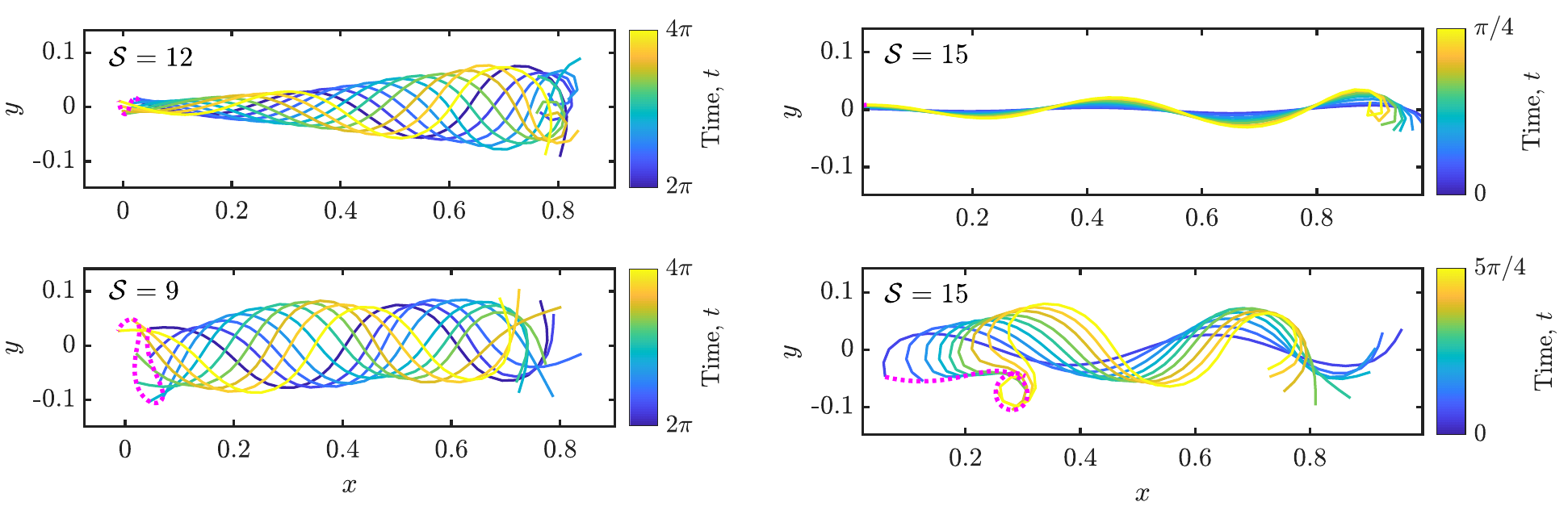}
	\caption{
		Shape configuration of single swimming filaments, propelled by the moment density profiles in Eqs.~\eqref{eq:sperm-beat} (top row) and \eqref{eq:worm-beat} (bottom row). In both scenarios, $ \mathfrak{m}_0 = 0.03 $, $ k=4\pi $, and $ Q=40 $. The dotted magenta curve traces the path of the leading point $ \bm{X}_0(t) $. On the left, an optimal choice of $ \mathcal{S} $ results in fast directed motion, while on the right, extreme filament curvature is produced, leading to self-intersection.
	}
	\label{figure10}
\end{figure}

Swimming speed for different choices of parameter pairs $ \left(\mathcal{S},\mathfrak{m}_0\right) $ are presented in Fig.~\ref{figure9}. For critical values, filaments self-intersect, in which case the EIF is inapplicable. The shape of such filaments are shown in Fig.~\ref{figure10}. For a sperm-like swimmer (left panel of Fig.~\ref{figure9}), swimming speed increases as $ \mathfrak{m}_0 $ and $ \mathcal{S} $ are increased in tandem for a fixed wave-number $ k=4\pi $. This is in contrast to worm-like swimmers (right panel of Fig.~\ref{figure9}), in which there is a clear optimal choice of $ \mathfrak{m}_0 $ for a given $ \mathcal{S} $ to induce fastest swimming. Shape profiles for the fastest swimmers of each swimming type are also displayed in Fig.~\ref{figure10}.

\subsection{Results: multiple passive filaments in shear flow} \label{subsec:results-multi-fil-shear}

\begin{figure}
	\centering
	\subfloat{\includegraphics[width=\textwidth]{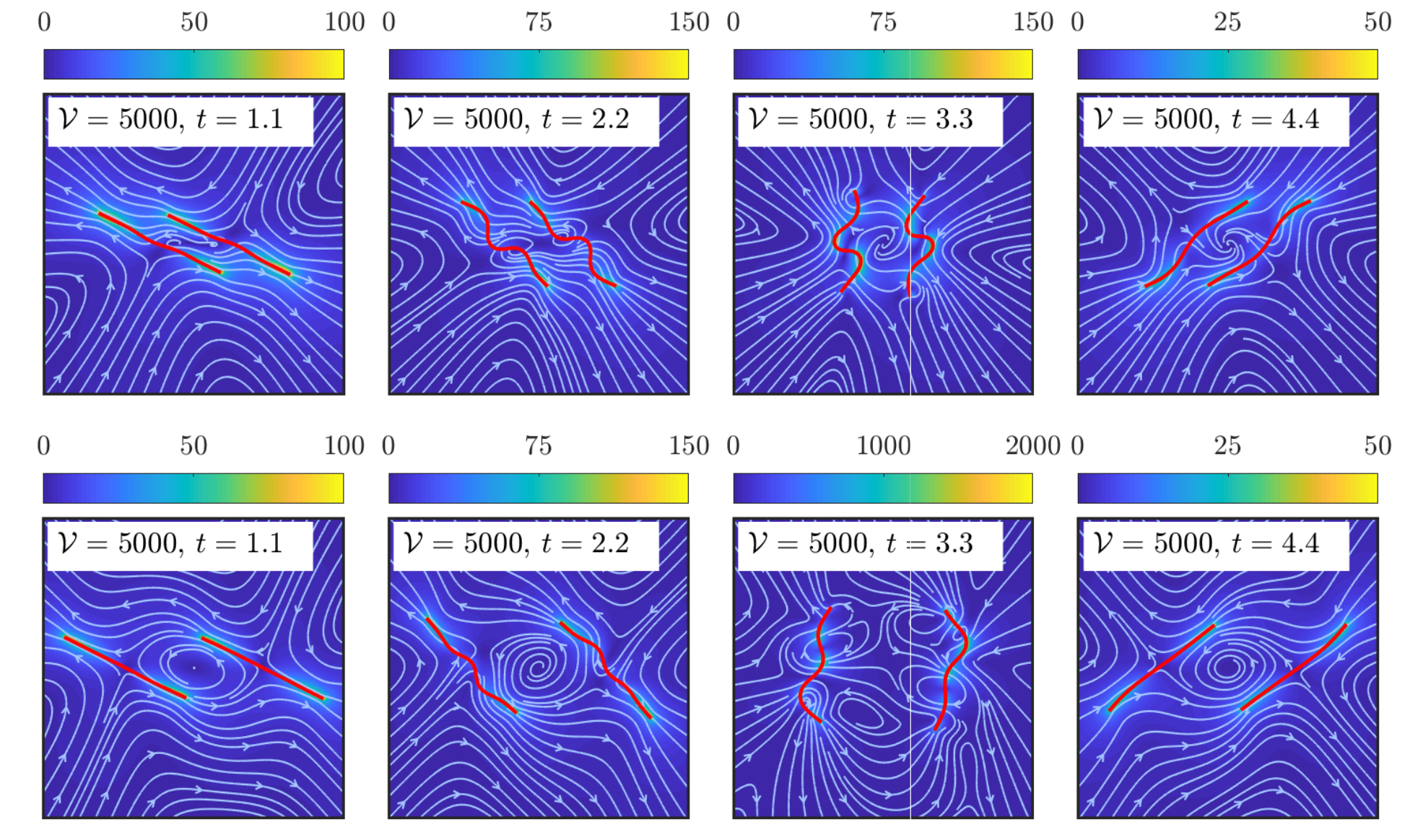} \label{figure11a}}
	\subfloat{\label{figure11b}} \\
	\caption{
		Dynamics of two identical filaments (with $ Q=40 $), separated by $ \Delta X_0=0.5 $ (top row), and $ \Delta X_0=1 $ (bottom row). The colorbar indicates the magnitude of the relative perturbed fluid velocity $\bm{U}_p = (\bm{U}_f - \bm{U}_b) / \bm{U}_{\text{ref}}$, where $\bm{U}_f$ is the fluid flow, $\bm{U}_b $ is the background fluid flow, and $\bm{U}_{\text{ref}}$ is the flow at an arbitrary reference field point, chosen to be the at the bottom-left of each figure frame. The plotted fluid lines display the flow disturbance, illustrating the filament-fluid interactions and the occlussionary effect that the presence of the left-most filament has on the flow.
	}
	\label{figure11}
\end{figure}
\begin{figure}
	\centering
	\includegraphics[width=\textwidth]{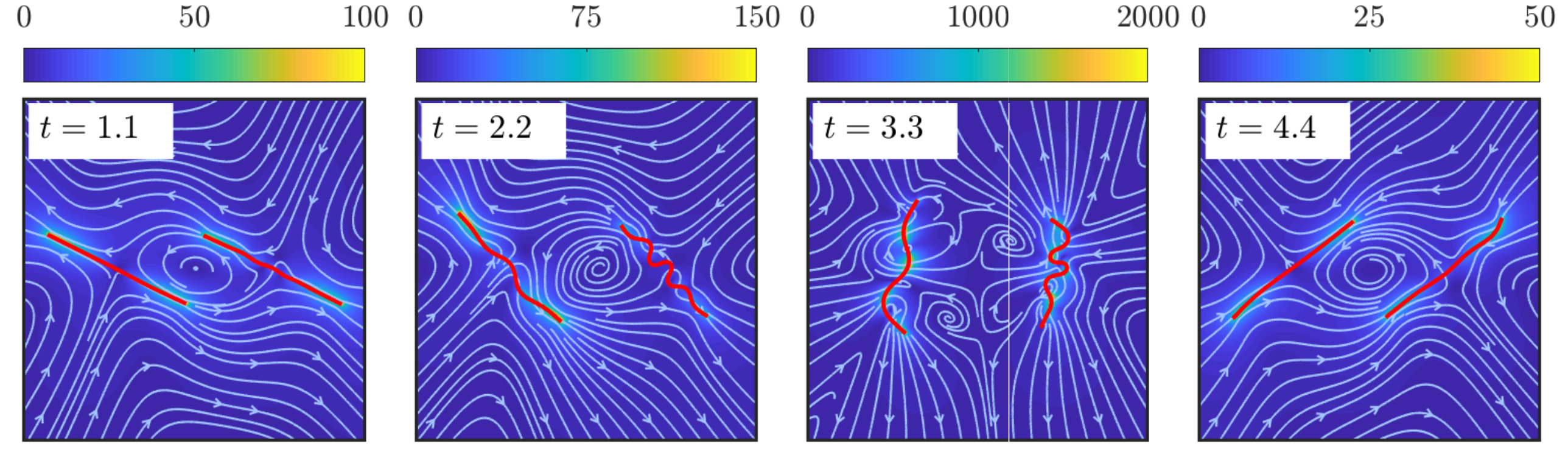} \\
	\caption{
		Two filaments of differing stiffness in shear flow, separated by $ \Delta X_0 = 1 $ and with $ Q=40 $. The left filament is characterized by $\mathcal{V}=5\times 10^3$, and the right by $\mathcal{V}=2\times 10^4$. The colorbar indicates the magnitude of the relative perturbed fluid velocity $\bm{U}_p = (\bm{U}_f - \bm{U}_b) / \bm{U}_{\text{ref}}$, where $\bm{U}_f$ is the fluid flow, $\bm{U}_b $ is the background fluid flow, and $\bm{U}_{\text{ref}}$ is the flow at an arbitrary reference field point, chosen to be the at the bottom-left of each figure frame.
	}
	\label{figure12}
\end{figure}
\begin{figure}[t]
    \centering
    \includegraphics[width=0.9\textwidth]{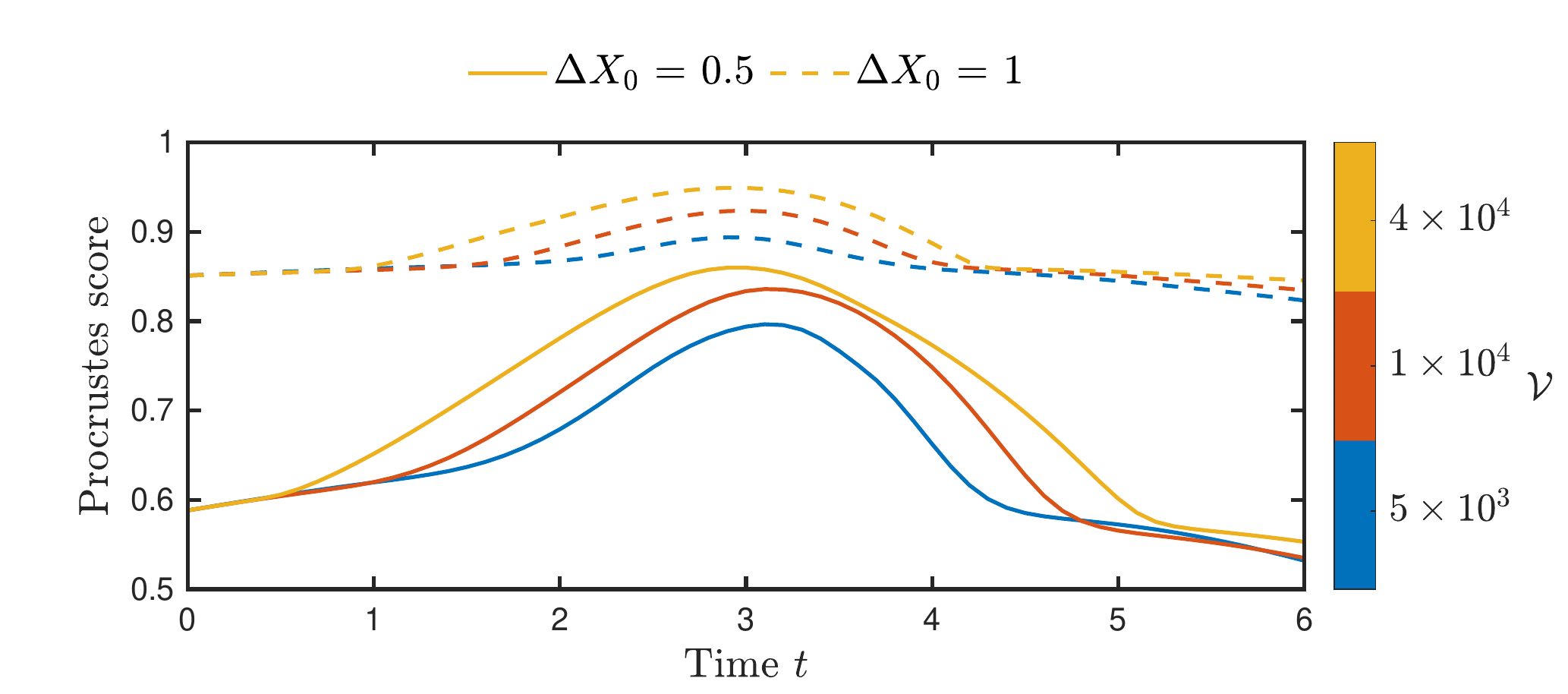}
    \caption{
    	Quantifying the evolving geometric similarity between two filaments of equal stiffness in shear flow for three values of $ \mathcal{V} $. The shape difference between two filaments is quantified using the Procrustes measure, calculated using the \texttt{procrustes} MATLAB\textsuperscript{\textregistered} command. Higher values indicate a larger degree of dissimilarity between the two filaments \cite{gower1975generalized}.
    }
    \label{figure13}
\end{figure}

Motivated by the work of Young \cite{young2009hydrodynamic}, two filaments of equal length are placed into a linear shear flow so that their initial midpoints ${\bm{X}(s=1/2,t=0)}$ intersect the line $y=0$, and are separated by a distance $ \Delta X_0 $. The initial shape profiles for each filament are the same as in Sec.~\ref{subsec:results-single-fil-shear}. For two filaments characterized by the same viscous-elastic parameter, non-local hydrodynamic interactions lead to geometric asymmetry, as seen in Fig.~\ref{figure11}. The value of $\mathcal{V}$ determines the level of dissimilarity between filaments with identical initial shape profiles, with higher values leading to larger deviations in shape throughout the rotation.

The initial separation distance also has a significant influence on the dynamics of each filament. The shapes of initially close filaments evolve in tandem, assuming similar geometries at a given moment in time (row one of Fig.~\ref{figure11}). Increasing the separation distance results in a decoupling of the filament shape profiles (row two of Fig.~\ref{figure11}). The EIF allows for each filament's characteristic parameter to be independently chosen, and in Fig.~\ref{figure12}, we highlight the variation in shape this can induce. In each of the panels in Figs.~\ref{figure11} and \ref{figure12}, color denotes the magnitude of the relative fluid velocity disturbance (found by subtracting the shear fluid velocity from the resultant fluid velocity), scaled with respect to the size of fluid velocity in the bottom left corner of each frame. Fluid lines show the instantaneous direction of the flow disturbance.

The geometric similarity between filaments can be assessed via their Procrustes score \cite{gower1975generalized}, calculated using the \texttt{procrustes} MATLAB\textsuperscript{\textregistered} command. We compare pairs of filaments initially separated by $\Delta X_0 = 0.5$ and $1$ for a range of $\mathcal{V}$, with results displayed in Fig.~\ref{figure13}. As $\mathcal{V}$ is increased, hydrodynamic interactions through the fluid cause larger levels of asymmetry between the filaments during the period of maximum buckling (when they are perpendicular to the direction of shear at $ t\approx3 $). Increasing the initial filament separation results in a higher baseline Procrustes score, but with reduced variability, demonstrating how non-local hydrodynamic interactions decay as the filaments are moved apart.

\subsection{Results: multiple passive filaments sedimenting under gravity}\label{subsec:results-multi-fil-sedimenting}

\begin{figure}
    \centering
    \includegraphics[width=\linewidth]{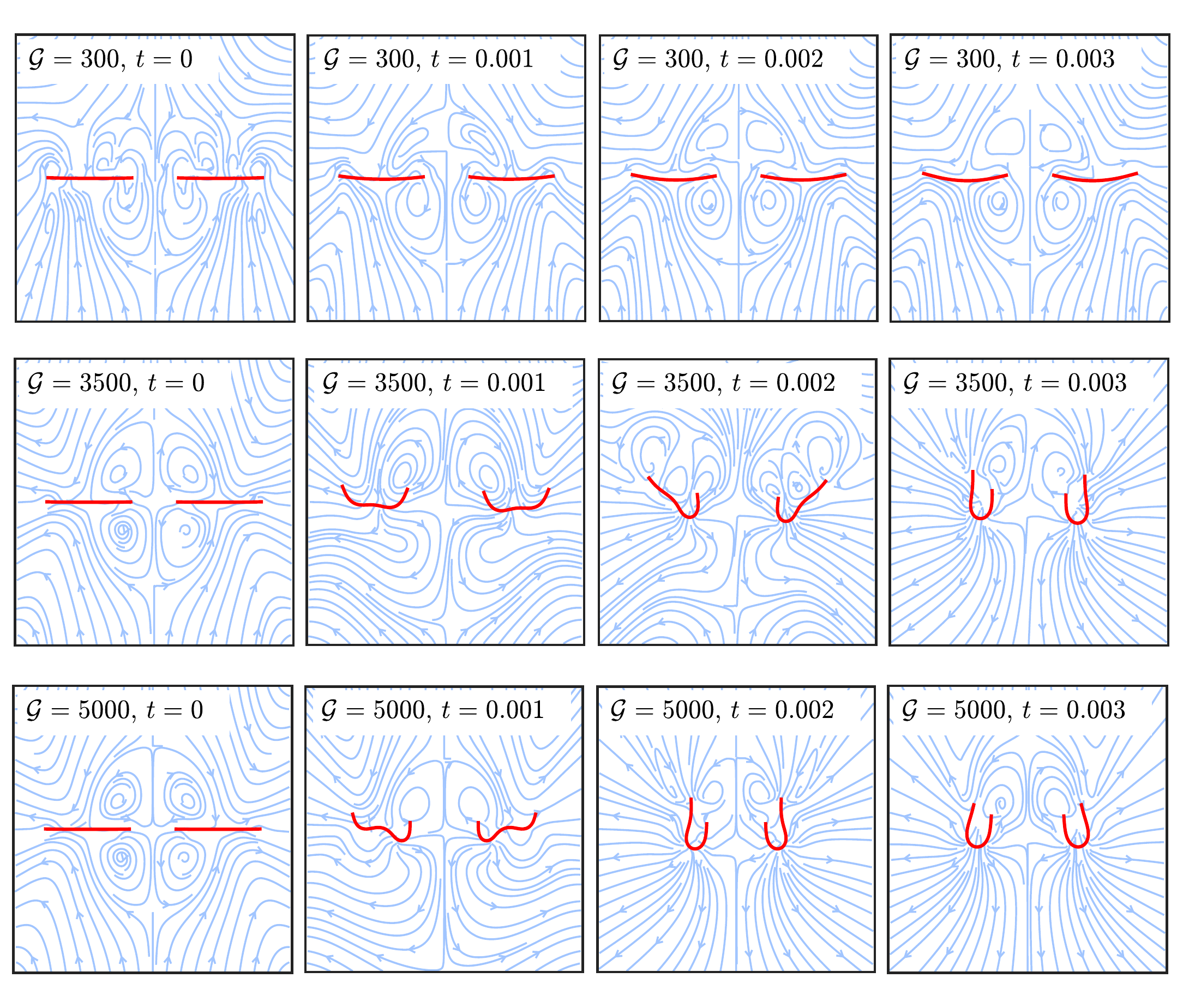}
    \caption{
    	Two filaments sedimenting under gravity, both characterized by the same elasto-gravitational parameter $\mathcal{G}$. Streamlines indicate the direction of the fluid disturbance caused by filament interactions.
    }
    \label{figure14}
\end{figure}

\begin{figure}
    \centering
    \includegraphics[width=1\textwidth]{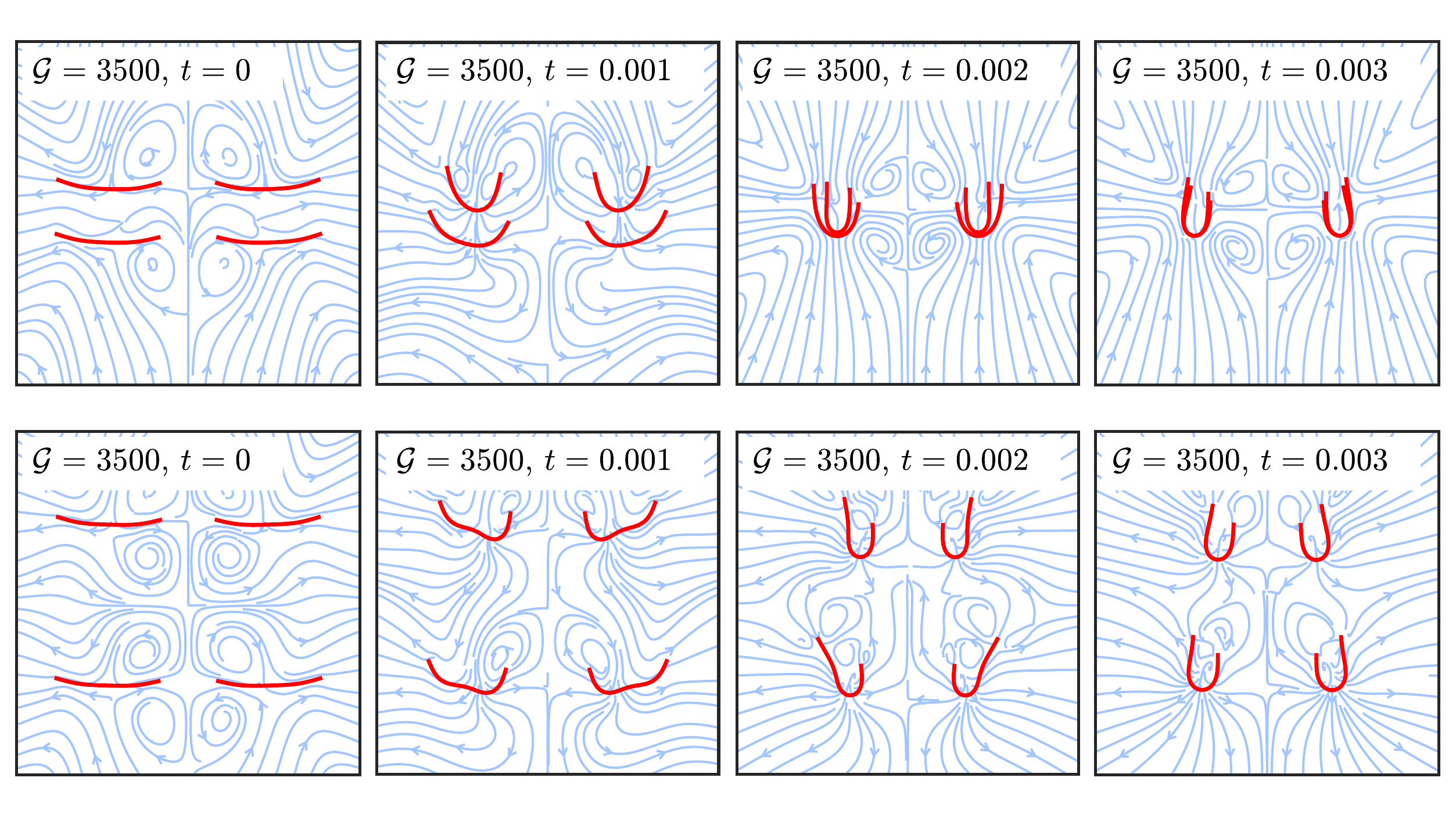}
    \caption{
    	The effect of initial filament density on group dynamics when sedimenting under gravity. In both experiments, the horizontal spacing between the filament columns are equal. A change in the vertical spacing from 0.5 (top row) to 1.5 (bottom row) leads to different fluid dynamics and resulting filament configurations. Streamlines indicate the direction of the fluid disturbance caused by filament interactions.
    }
    \label{figure15}
\end{figure}

\begin{figure}
    \centering
    \includegraphics[width=1\textwidth]{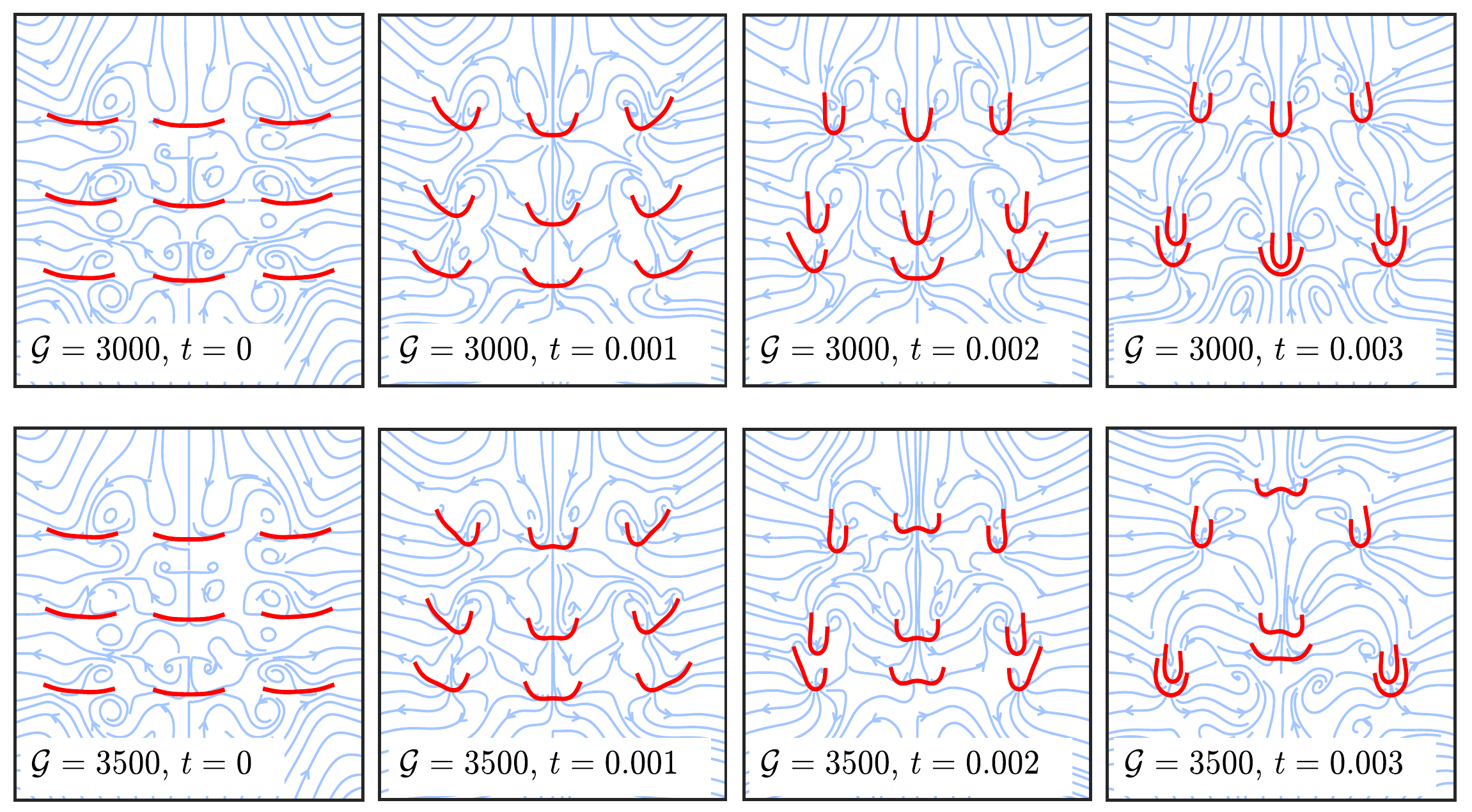}
    \caption{
    	Simulation of multiple sedimenting filaments. The EIF framework can also accommodate larger arrays of filaments, such as the $ 3\times 3 $ arrangements in the above figure. In each case, the initial filament spacing is $\Delta \bm{X}_0 = [1.5,1] $. In all cases, vertical symmetry breaking is apparent for $t>0.001$. Streamlines indicate the direction of the fluid disturbance caused by filament interactions.
    }
    \label{figure16}
\end{figure}

A range of filament systems with multiple choices of the elasto-gravitational parameter $\mathcal{G}$ are presented in Figs.~\ref{figure14}, \ref{figure15} and \ref{figure16}. In each case, the initial filament shape configurations are as those in Sec.~\ref{subsec:results-single-fil-sedimenting}, and are separated from each other by $ \Delta \bm{X}_0 = [\Delta X_0,\Delta Y_0]$. A stopping criterion is implemented which halts the integration when filaments intersect or otherwise touch. For small values of $ \mathcal{G} $ filaments slide towards each other as they sediment due to the anisotropy of Stokes drag. For larger values of $\mathcal{G}$ (rows two and three of Fig.~\ref{figure14}) the metastable and stable buckling modes observed in the single filament experiments (Figs.~\ref{figure7} and \ref{figure8}) are replicated. In these systems, the interaction of multiple filaments leads to the steady-state profiles being reached sooner than in the single-filament cases. Furthermore, the onset of buckling occurs at reduced values of $\mathcal{G}$ when multiple filaments are interacting.

As was the case for multiple filaments in shear flow, the inter-filament spacing $ \Delta \bm{X}_0 $ also has a significant effect on the resulting group dynamics, and can lead to symmetry breaking in the arrangement of filaments. In Fig.~\ref{figure15}, two initial filament placement configurations are tested. For each setup, we initialize four filaments arranged in a grid, with $\Delta X_0 = 1.5$ and vertical separations $\Delta Y_0 = 0.5$ (row one of Fig.~\ref{figure15}) and $\Delta Y_0 = 1.5$ (row two of Fig.~\ref{figure15}). Smaller vertical spacing leads to the filaments ``nestling'' in a horizontally-mirrored configuration; when placed further apart, horizontal mirroring still occurs, but the filaments do not approach each other.

In Fig.~\ref{figure16} nine identical filaments are arranged in a grid with initial spacing ${\Delta X_0 = 1.5}$ and $\Delta Y_0 = 1$. As with the smaller arrays, filaments can group and buckle according to the choice of characteristic parameter $\mathcal{G}$. Although identical in governing parameter $ \mathcal{G} $, the non-local hydrodynamic interactions between filaments induce different buckling behaviors depending on location within the array.

For all choices of $\mathcal{G}$, vertical symmetry-breaking occurs between the top-most and middle rows of filaments, with the second and third rows of filaments on the left and right of the frame nestling into those below them. For $\mathcal{G}=3500$, more prominent buckling is apparent, with filaments in the central column approaching a ``W'' shape, whereas those on the flanks assume a ``U'' configuration. Competing interactions between filaments in the central column and their surrounding neighbors causes them to remain in a metastable ``W'' shape longer than would be expected in the single-filament case.

\subsection{Results: multiple active filaments} \label{subsec:results-multi-fil-active}

\begin{figure}
	\centering
	\includegraphics[width=0.8\linewidth]{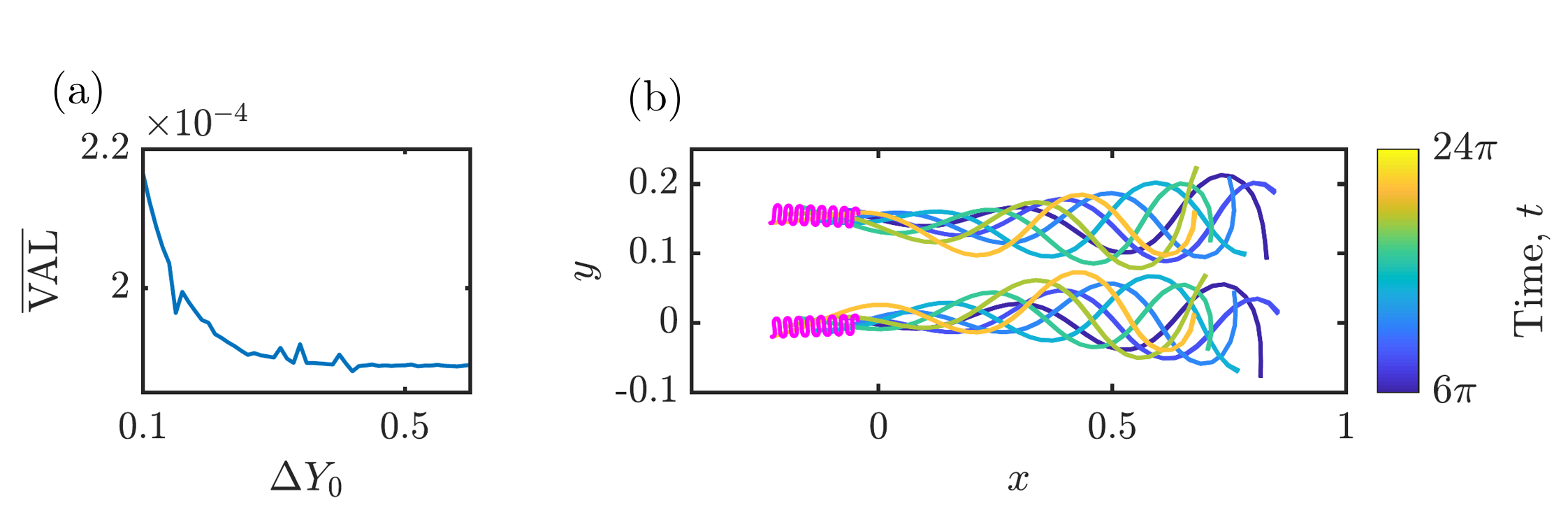} \\ \vspace{-0.3cm}
	\caption{
		Two identical filaments with \(\mathcal{S}=8\) and \(\mathfrak{m}_0=0.05\), propelled by a sperm-like active moment, swimming alongside one-another. In (a), filaments in close proximity of each other produce a higher average velocity ($ \overline{\textup{VAL}} $). In (b), the shapes of the evolving filaments are presented for an initial vertical separation $ \Delta Y_0 =0.1$ for $ t\in[6\pi,24\pi] $. As in Fig.~\ref{figure10}, the magenta paths in (b) indicate the path the leading point of the filament takes over the course of the simulation.
	}
\label{figure17}
\end{figure}

Multiple self-propulsive active filaments can  be simulated using the EIF method. To illustrate this, we consider two filaments swimming alongside each other at different separation distances $ \Delta Y_0 $, with shapes initialized as in Sec.~\ref{subsec:single-fil-swimming}. For fixed $ \mathcal{S}=8 $ and $ \mathfrak{m}_0 = 0.05 $, the size of the separation determines the resulting average swimming velocity, as shown in Fig.~\ref{figure17}. At low levels of separation, hydrodynamic interactions between the synchronous beats result in a higher average VAL (Eq.~\eqref{def:VAL}), which decays as the filaments are placed further apart (Fig.~\ref{figure17}a).

\section{Discussion} \label{sec:discussion}

In this manuscript, the elastohydrodynamic integral equation (EIF) framework is formulated, and applied to problems involving single and multiple filaments in shear flows, sedimenting under gravity, and swimming due to a prescribed active moment. From the simulations presented in Sec.~\ref{sec:results}, it is apparent how inter-filament non-local hydrodynamic interactions have a role in governing filament shapes and buckling behavior. By examining active moment driven swimmers, an optimum pairing of moment-amplitude and characteristic swimming number is revealed.

One of the key benefits to the integral formulation framework presented is that the need for computing Lagrange multipliers of tension is removed; inextensibility of the modeled filament is already guaranteed by the use of the method of lines discretization. This results in a method that is more computationally efficient than many similar contemporary models. The EIF method proposed by Moreau and co-authors \cite{moreau2018asymptotic} highlighted the reduction in computational runtime attainable when using the integral formulation applied to elastodynamics, due in part to the reduction in the required degrees of freedom the integral formulation affords. Alleviating this numerical stiffness facilitates the study of non-linear problems, such as those involving filament buckling investigated in Sec.~\ref{sec:results}. Additional computational costs are incurred by introducing non-local hydrodynamics to the EIF framework, in line with the increase in the dimension and density of the matrix system solved at each time-step. The BLM and the EIF with regularized stokeslets complete a relaxing rod simulation in about equivalent walltime (as expected, since both approaches compute three-dimensional non-local hydrodynamics), shown in Fig.~\ref{figure2c}. However, to ensure that the point-wise error is small, fine discretizations need to be used with the bead model ($Q>60$), whereas the EIF method performs equally well with a much coarser segmentation of the filament ($Q>20$). This allows for adequately accurate results to be obtained for reduced computational cost when using the EIF with regularized stokeslets over the bead and link model.

The proposed method has the potential to quickly and accurately simulate arrays of filaments in various flows and surroundings. The EIF method presented here will enable the solution of more challenging problems such as planar beating cilia sheets, or multiple sperm swimming in a narrow channel. The methods' modular framework enables such problems to be setup and executed in a simple and obvious manner. Additionally, the EIF method fits into the family of regularized stokeslet methods \cite{cortez2005method}, which are increasingly widely used for problems in biological fluid mechanics. Crucially, the computational efficiency exhibited in the single filament experiments is retained. The method could be extended by incorporating the treecode formulation of Wang \textit{et al.} \cite{wang2018treecode}, or equivalent methods, to coarse-grain the far-field flow, simplifying the computational cost and enabling simulation of increasingly large numbers of filaments. The inclusion of a repulsive force term, such as the Lennard-Jones potential employed by Jarayaman \textit{et al.} \cite{jayaraman2012autonomous}, could additionally enable simulation of filaments in close proximity and help avoid filament self-intersection.

The results of Sec.~\ref{subsec:results-single-fil-shear} suggest that the shape of a buckling filament can be described to a high degree of accuracy by a relatively low order Chebyshev polynomial (Fig.~\ref{figure6}). These results suggest that a modified discretization based on orthogonal polynomials, perhaps in concert with suitable quadrature techniques \cite{muldowney1995spectral} might provide further improvements in efficiency and scalability. In summary we hope that the integral operator formulation of elastohydrodynamics will be valuable in biological fluid dynamics and beyond.

\section{Acknowledgments}

D.J.S. and M.T.G. acknowledge funding from the Engineering and Physical Sciences Research Council (EPSRC), Healthcare Technologies Challenge Award (EP/N021096/1). T.D.M-J. acknowledges funding from the EPSRC (EP/R041555/1). A.L.H-M. acknowledges support from the EPSRC for funding via a PhD scholarship (EP/N509590/1).


\begin{thebibliography}{10}

\bibitem{moreau2018asymptotic}
C.~Moreau, L.~Giraldi, and H.~Gad{\^e}lha.
\newblock The asymptotic coarse-graining formulation of slender-rods,
  bio-filaments and flagella.
\newblock {\em J. R. Soc., Interface}, {\bf 15}(144):20180235, 2018.

\bibitem{schoeller2019method}
S.{\,}F. Schoeller, A.{\,}K. Townsend, T.{\,}A. Westwood, and E.{\,}E. Keaveny.
\newblock Methods for suspensions of passive and active filaments.
\newblock {\em arXiv preprint arXiv:1903.12609}, 2019.

\bibitem{jayaraman2012autonomous}
G.~Jayaraman, S.~Ramachandran, S.~Ghose, A.~Laskar, M.{\,}S. Bhamla,
  P.{\,}B.~Sunil Kumar, and R.~Adhikari.
\newblock Autonomous motility of active filaments due to spontaneous
  flow-symmetry breaking.
\newblock {\em Phys. Rev. Lett.}, {\bf 109}(15):158302, 2012.

\bibitem{delmotte2015general}
B.~Delmotte, E.~Climent, and F.~Plourabou{\'e}.
\newblock A general formulation of {B}ead {M}odels applied to flexible fibers
  and active filaments at low {R}eynolds number.
\newblock {\em J. Comput. Phys.}, {\bf 286}:14--37, 2015.

\bibitem{brokaw1971bend}
C.{\,}J. Brokaw.
\newblock Bend propagation by a sliding filament model for flagella.
\newblock {\em J. Exp. Biol.}, {\bf 55}(2):289--304, 1971.

\bibitem{brokaw1972computer}
C.{\,}J. Brokaw.
\newblock Computer simulation of flagellar movement: {I}. {D}emonstration of
  stable bend propagation and bend initiation by the sliding filament model.
\newblock {\em Biophys. J.}, {\bf 12}(5):564--586, 1972.

\bibitem{hines1978bend}
M.~Hines and J.{\,}J. Blum.
\newblock Bend propagation in flagella. {I}. {D}erivation of equations of
  motion and their simulation.
\newblock {\em Biophys. J.}, {\bf 23}(1):41--57, 1978.

\bibitem{gadelha2010nonlinear}
H.~Gad{\^e}lha, E.{\,}A. Gaffney, D.{\,}J. Smith, and J.{\,}C. Kirkman-Brown.
\newblock Nonlinear instability in flagellar dynamics: a novel modulation
  mechanism in sperm migration?
\newblock {\em J. R. Soc., Interface}, {\bf 7}(53):1689--1697, 2010.

\bibitem{tornberg2004simulating}
A.-K. Tornberg and M.{\,}J. Shelley.
\newblock Simulating the dynamics and interactions of flexible fibers in
  {S}tokes flows.
\newblock {\em J. Comput. Phys.}, {\bf 196}(1):8--40, 2004.

\bibitem{cortez2018regularized}
R.~Cortez.
\newblock Regularized {S}tokeslet segments.
\newblock {\em J. Comput. Phys.}, {\bf 375}:783--796, 2018.

\bibitem{gallagher2018meshfree}
M.{\,}T. Gallagher and D.{\,}J. Smith.
\newblock Meshfree and efficient modeling of swimming cells.
\newblock {\em Phys. Rev. Fluids}, {\bf 3}(5):053101, 2018.

\bibitem{lindemann1994model}
C.{\,}B. Lindemann.
\newblock A model of flagellar and ciliary functioning which uses the forces
  transverse to the axoneme as the regulator of dynein activation.
\newblock {\em Cell Motil. Cytoskel.}, {\bf 29}(2):141--154, 1994.

\bibitem{guo2018bistability}
H.~Guo, L.~Fauci, M.~Shelley, and E.~Kanso.
\newblock Bistability in the synchronization of actuated microfilaments.
\newblock {\em J. Fluid Mech.}, {\bf 836}:304--323, 2018.

\bibitem{oriola2017nonlinear}
D.~Oriola, H.~Gad{\^e}lha, and J.~Casademunt.
\newblock Nonlinear amplitude dynamics in flagellar beating.
\newblock {\em R. Soc. Open Sci.}, {\bf 4}(3):160698, 2017.

\bibitem{huang2018hydrodynamic}
J.~Huang, L.~Carichino, and S.{\,}D. Olson.
\newblock Hydrodynamic interactions of actuated elastic filaments near a planar
  wall with applications to sperm motility.
\newblock {\em J. Coupled Syst. Multiscale Dyn.}, {\bf 6}(3):163--175, 2018.

\bibitem{gadelha2013counterbend}
H.~Gad{\^e}lha, E.{\,}A. Gaffney, and A.~Goriely.
\newblock The counterbend phenomenon in flagellar axonemes and cross-linked
  filament bundles.
\newblock {\em Proc. Natl. Acad. Sci.}, {\bf 110}(30):12180--12185, 2013.

\bibitem{smith2009bend}
D.{\,}J. Smith, E.{\,}A. Gaffney, H.~Gad{\^e}lha, N.~Kapur, and J.{\,}C.
  Kirkman-Brown.
\newblock Bend propagation in the flagella of migrating human sperm, and its
  modulation by viscosity.
\newblock {\em Cell Motil. Cytoskeleton}, {\bf 66}(4):220--236, 2009.

\bibitem{montenegro2015spermatozoa}
T.{\,}D. Montenegro-Johnson, H.~Gadelha, and D.{\,}J. Smith.
\newblock Spermatozoa scattering by a microchannel feature: an
  elastohydrodynamic model.
\newblock {\em R. Soc. Open Sci.}, {\bf 2}(3):140475, 2015.

\bibitem{denissenko2012human}
P.~Denissenko, V.~Kantsler, D.{\,}J. Smith, and J.~Kirkman-Brown.
\newblock Human spermatozoa migration in microchannels reveals
  boundary-following navigation.
\newblock {\em Proc. Natl. Acad. Sci.}, {\bf 109}(21):8007--8010, 2012.

\bibitem{woolley2001study}
D.{\,}M. Woolley and G.{\,}G. Vernon.
\newblock A study of helical and planar waves on sea urchin sperm flagella,
  with a theory of how they are generated.
\newblock {\em J. Exp. Biol.}, {\bf 204}(7):1333--1345, 2001.

\bibitem{gallagher2018casa}
M.{\,}T. Gallagher, D.{\,}J. Smith, and J.{\,}C. Kirkman-Brown.
\newblock {CASA}: {T}racking the past and plotting the future.
\newblock {\em Reprod. Fert. Develop.}, {\bf 30}(6):867--874, 2018.

\bibitem{montenegro2012modelling}
T.{\,}D. Montenegro-Johnson, A.{\,}A. Smith, D.{\,}J. Smith, D.~Loghin, and
  J.{\,}R. Blake.
\newblock Modelling the fluid mechanics of cilia and flagella in reproduction
  and development.
\newblock {\em Eur. Phys. J. E}, {\bf 35}(10):111, 2012.

\bibitem{ferreira2017physical}
R.{\,}R. Ferreira, A.~Vilfan, F.~J{\"u}licher, W.~Supatto, and J.~Vermot.
\newblock Physical limits of flow sensing in the left-right organizer.
\newblock {\em eLife}, {\bf 6}:e25078, 2017.

\bibitem{omori2017nodal}
T.~Omori, H.~Sugai, Y.~Imai, and T.~Ishikawa.
\newblock Nodal cilia-driven flow: development of a computational model of the
  nodal cilia axoneme.
\newblock {\em J. Biomech.}, {\bf 61}:242--249, 2017.

\bibitem{gadelha2013optimal}
H.~Gad{\^e}lha.
\newblock On the optimal shape of magnetic swimmers.
\newblock {\em Regul. Chaotic Dyn.}, {\bf 18}(1-2):75--84, 2013.

\bibitem{pacheco2011detection}
R.~Pacheco-G{\'o}mez, J.~Kraemer, S.~Stokoe, H.{\,}J. England, C.{\,}W. Penn,
  E.~Stanley, A.~Rodger, J.~Ward, M.{\,}R. Hicks, and T.{\,}R. Dafforn.
\newblock Detection of pathogenic bacteria using a homogeneous immunoassay
  based on shear alignment of virus particles and linear dichroism.
\newblock {\em Anal. Chem.}, {\bf 84}(1):91--97, 2011.

\bibitem{lobo2015direct}
D.{\,}P. Lobo, A.{\,}M. Wemyss, D.{\,}J. Smith, A.~Straube, K.{\,}B.
  Betteridge, A.{\,}H.{\,}J. Salmon, R.{\,}R. Foster, H.{\,}E. Elhegni,
  S.{\,}C. Satchell, H.{\,}A. Little, et~al.
\newblock Direct detection and measurement of wall shear stress using a
  filamentous bio-nanoparticle.
\newblock {\em Nano Res.}, {\bf 8}(10):3307--3315, 2015.

\bibitem{gallagher2017model}
M.{\,}T. Gallagher, C.{\,}V. Neal, K.{\,}P. Arkill, and D.{\,}J. Smith.
\newblock Model-based image analysis of a tethered {B}rownian fibre for shear
  stress sensing.
\newblock {\em J. R. Soc., Interface}, {\bf 14}(137):20170564, 2017.

\bibitem{montenegro2018microtransformers}
T.{\,}D. Montenegro-Johnson.
\newblock Microtransformers: {C}ontrolled microscale navigation with flexible
  robots.
\newblock {\em Phys. Rev. Fluids}, {\bf 3}(6):062201, 2018.

\bibitem{coy2017counterbend}
R.~Coy and H.~Gad{\^e}lha.
\newblock The counterbend dynamics of cross-linked filament bundles and
  flagella.
\newblock {\em J. R. Soc., Interface}, {\bf 14}(130):20170065, 2017.

\bibitem{yang2017dynamics}
Q.~Yang and L.~Fauci.
\newblock Dynamics of a macroscopic elastic fibre in a polymeric cellular flow.
\newblock {\em J. Fluid Mech.}, {\bf 817}:388--405, 2017.

\bibitem{wrobel2016enhanced}
J.{\,}K. Wr{\'o}bel, S.~Lynch, A.~Barrett, L.~Fauci, and R.~Cortez.
\newblock Enhanced flagellar swimming through a compliant viscoelastic network
  in stokes flow.
\newblock {\em J. Fluid Mech.}, {\bf 792}:775--797, 2016.

\bibitem{cortez2001method}
R.~Cortez.
\newblock The method of regularized {S}tokeslets.
\newblock {\em SIAM J. Sci. Comput.}, {\bf 23}(4):1204--1225, 2001.

\bibitem{cortez2005method}
R.~Cortez, L.~Fauci, and A.~Medovikov.
\newblock The method of regularized {S}tokeslets in three dimensions: analysis,
  validation, and application to helical swimming.
\newblock {\em Phys. Fluids}, {\bf 17}(3):031504, 2005.

\bibitem{stein2019coarse}
D.{\,}B. Stein and M.{\,}J. Shelley.
\newblock Coarse-graining the dynamics of immersed and driven fiber assemblies.
\newblock {\em arXiv preprint arXiv:1902.00049}, 2019.

\bibitem{olson2015hydrodynamic}
S.{\,}D. Olson and L.{\,}J. Fauci.
\newblock Hydrodynamic interactions of sheets vs filaments: {S}ynchronization,
  attraction, and alignment.
\newblock {\em Phys. Fluids}, {\bf 27}(12):121901, 2015.

\bibitem{martin2019use}
P.{\,}A. Martin.
\newblock On the use of approximate fundamental solutions: {C}onnections with
  the method of fundamental solutions and the method of regularized stokeslets.
\newblock {\em Eng. Anal. Boundary Elem.}, {\bf 99}:23--28, 2019.

\bibitem{smith2009boundary}
D.{\,}J. Smith.
\newblock A boundary element regularized {S}tokeslet method applied to
  cilia-and flagella-driven flow.
\newblock {\em Proc. R. Soc. A}, {\bf 465}(2112):3605--3626, 2009.

\bibitem{shampine1997matlab}
L.{\,}F. Shampine and M.{\,}W. Reichelt.
\newblock The {MATLAB} {ODE} suite.
\newblock {\em SIAM J. Sci. Comput.}, {\bf 18}(1):1--22, 1997.

\bibitem{liu2018morphological}
Y.~Liu, B.~Chakrabarti, D.~Saintillan, A.~Lindner, and O.~du~Roure.
\newblock Morphological transitions of elastic filaments in shear flow.
\newblock {\em Proc. Natl. Acad. Sci.}, {\bf 115}(38):9438--9443, 2018.

\bibitem{young2009hydrodynamic}
Y.-N. Young.
\newblock Hydrodynamic interactions between two semiflexible inextensible
  filaments in {S}tokes flow.
\newblock {\em Phys. Rev. E}, {\bf 79}(4):046317, 2009.

\bibitem{becker2001instability}
L.{\,}E. Becker and M.{\,}J. Shelley.
\newblock Instability of elastic filaments in shear flow yields
  first-normal-stress differences.
\newblock {\em Phys. Rev. Lett.}, {\bf 87}(19):198301, 2001.

\bibitem{consentino2005hydrodynamic}
M.{\,}C. Lagomarsino, I.~Pagonabarraga, and C.{\,}P. Lowe.
\newblock Hydrodynamic induced deformation and orientation of a microscopic
  elastic filament.
\newblock {\em Phys. Rev. Lett.}, {\bf 94}(14):148104, 2005.

\bibitem{gower1975generalized}
J.{\,}C. Gower.
\newblock Generalized procrustes analysis.
\newblock {\em Psychometrika}, {\bf 40}(1):33--51, 1975.

\bibitem{wang2018treecode}
L.~Wang, S.~Tlupova, and R.~Krasny.
\newblock A {T}reecode {A}lgorithm for {3D} {S}tokeslets and {S}tresslets.
\newblock {\em Adv. Appl. Math. Mech}, {\bf 11}(4):737--756, 2019.

\bibitem{muldowney1995spectral}
G.{\,}P. Muldowney and J.{\,}J.{\,}L. Higdon.
\newblock A spectral boundary element approach to three-dimensional {S}tokes
  flow.
\newblock {\em J. Fluid Mech.}, {\bf 298}:167--192, 1995.

\end{thebibliography}
\end{document}